\let\accentvec\vec     
\documentclass[pdftex,twocolumn, epjc3]{svjour3}          % twocolumn
\RequirePackage[T1]{fontenc}

\smartqed  % flush right qed marks, e.g. at end of proof

\RequirePackage{graphicx}
\usepackage[british]{babel}
\usepackage{ifthen} % for conditional statements
\usepackage{xspace}
\let\vec\accentvec
\usepackage{comment}
\usepackage{amsmath} % Adds a large collection of math symbols
\usepackage{amssymb}
\usepackage{amsfonts}
\usepackage{upgreek} % Adds in support for greek letters in roman typeset
\graphicspath{{./figs/}} % Make Latex search fig subdir for figures

\RequirePackage[colorlinks,citecolor=blue,urlcolor=blue,linkcolor=blue]{hyperref}
\usepackage[all]{hypcap} % Internal hyperlinks to floats.

\newboolean{articletitles}
\setboolean{articletitles}{true} % False removes titles in references

\newboolean{uprightparticles}
\setboolean{uprightparticles}{false} %True for upright particle symbols

\usepackage{xspace} 
\usepackage{upgreek}

\def\lhcb {\mbox{LHCb}\xspace}

\def\babar  {\mbox{BaBar}\xspace}
\def\belle  {\mbox{Belle}\xspace}

\def\MagUp {\mbox{\em Mag\kern -0.05em Up}\xspace}

\ifthenelse{\boolean{uprightparticles}}%
{

 \def\Pmu         {\ensuremath{\upmu}\xspace}                 
 \def\Pnu         {\ensuremath{\upnu}\xspace}                 
                  
 \def\Ppi         {\ensuremath{\uppi}\xspace}

 \def\Ptau        {\ensuremath{\uptau}\xspace}

 \def\PDelta      {\ensuremath{\Delta}\xspace}                 
 \def\PXi      {\ensuremath{\Xi}\xspace}                 
 \def\PLambda      {\ensuremath{\Lambda}\xspace}                 
 \def\PSigma      {\ensuremath{\Sigma}\xspace}                 
 \def\POmega      {\ensuremath{\Omega}\xspace}                 
 \def\PUpsilon      {\ensuremath{\Upsilon}\xspace}                 
 
 \def\PB      {\ensuremath{\mathrm{B}}\xspace}                 
                  
 \def\PD      {\ensuremath{\mathrm{D}}\xspace}

 \def\PK      {\ensuremath{\mathrm{K}}\xspace}

 \def\PW      {\ensuremath{\mathrm{W}}\xspace}

 \def\PZ      {\ensuremath{\mathrm{Z}}\xspace}                 
                  
 \def\Pb      {\ensuremath{\mathrm{b}}\xspace}                 
 \def\Pc      {\ensuremath{\mathrm{c}}\xspace}                 
                  
 \def\Pe      {\ensuremath{\mathrm{e}}\xspace}

 \def\Pi      {\ensuremath{\mathrm{i}}\xspace}

}
{

 \def\Pmu         {\ensuremath{\mu}\xspace}                 
 \def\Pnu         {\ensuremath{\nu}\xspace}                 
                  
 \def\Ppi         {\ensuremath{\pi}\xspace}

 \def\Ptau        {\ensuremath{\tau}\xspace}

 \mathchardef\PDelta="7101
 \mathchardef\PXi="7104
 \mathchardef\PLambda="7103
 \mathchardef\PSigma="7106
 \mathchardef\POmega="710A
 \mathchardef\PUpsilon="7107
                  
 \def\PB      {\ensuremath{B}\xspace}                 
                  
 \def\PD      {\ensuremath{D}\xspace}

 \def\PK      {\ensuremath{K}\xspace}

 \def\PW      {\ensuremath{W}\xspace}

 \def\PZ      {\ensuremath{Z}\xspace}                 
                  
 \def\Pb      {\ensuremath{b}\xspace}                 
 \def\Pc      {\ensuremath{c}\xspace}                 
                  
 \def\Pe      {\ensuremath{e}\xspace}

 \def\Pi      {\ensuremath{i}\xspace}

}

\makeatletter
\ifcase \@ptsize \relax% 10pt
  \newcommand{\miniscule}{\@setfontsize\miniscule{4}{5}}% \tiny: 5/6
\or% 11pt
  \newcommand{\miniscule}{\@setfontsize\miniscule{5}{6}}% \tiny: 6/7
\or% 12pt
  \newcommand{\miniscule}{\@setfontsize\miniscule{5}{6}}% \tiny: 6/7
\fi
\makeatother

\DeclareRobustCommand{\optbar}[1]{\shortstack{{\miniscule (\rule[.5ex]{1.25em}{.18mm})}
  \\ [-.7ex] $#1$}}

\def\en         {{\ensuremath{\Pe^-}}\xspace}   % electron negative (\em is taken)

\def\mun        {{\ensuremath{\Pmu^-}}\xspace} % muon negative (\mum is taken)

\def\taum       {{\ensuremath{\Ptau^-}}\xspace}

\def\ellm       {{\ensuremath{\ell^-}}\xspace}

\def\neu        {{\ensuremath{\Pnu}}\xspace}
\def\neub       {{\ensuremath{\overline{\Pnu}}}\xspace}

\def\neumb      {{\ensuremath{\neub_\mu}}\xspace}
\def\neut       {{\ensuremath{\neu_\tau}}\xspace}
\def\neutb      {{\ensuremath{\neub_\tau}}\xspace}
\def\neul       {{\ensuremath{\neu_\ell}}\xspace}
\def\neulb      {{\ensuremath{\neub_\ell}}\xspace}
\def\W      {{\ensuremath{\PW}}\xspace}

\def\Z      {{\ensuremath{\PZ}}\xspace}

\def\cquark    {{\ensuremath{\Pc}}\xspace}

\def\bquark    {{\ensuremath{\Pb}}\xspace}

\def\pion   {{\ensuremath{\Ppi}}\xspace}

\def\pip    {{\ensuremath{\pion^+}}\xspace}

\def\kaon    {{\ensuremath{\PK}}\xspace}
  \def\Kbar    {{\kern 0.2em\overline{\kern -0.2em \PK}{}}\xspace}

\def\KorKbar    {\kern 0.18em\optbar{\kern -0.18em K}{}\xspace}

\def\Km      {{\ensuremath{\kaon^-}}\xspace}

  \def\Dbar    {{\kern 0.2em\overline{\kern -0.2em \PD}{}}\xspace}
\def\D       {{\ensuremath{\PD}}\xspace}

\def\DorDbar    {\kern 0.18em\optbar{\kern -0.18em D}{}\xspace}
\def\Dz      {{\ensuremath{\D^0}}\xspace}

\def\Dp      {{\ensuremath{\D^+}}\xspace}

\def\Dstar   {{\ensuremath{\D^*}}\xspace}

\def\B       {{\ensuremath{\PB}}\xspace}
\def\Bbar    {{\ensuremath{\kern 0.18em\overline{\kern -0.18em \PB}{}}}\xspace}

\def\BorBbar    {\kern 0.18em\optbar{\kern -0.18em B}{}\xspace}

\def\Bzb     {{\ensuremath{\Bbar{}^0}}\xspace}

\def\Bub     {{\ensuremath{\B^-}}\xspace}

\def\Bm      {{\ensuremath{\Bub}}\xspace}

  \def\Y#1S{\ensuremath{\PUpsilon{(#1S)}}\xspace}% no space before {...}!

\def\Lbar        {{\ensuremath{\kern 0.1em\overline{\kern -0.1em\PLambda}}}\xspace}
\def\LorLbar    {\kern 0.18em\optbar{\kern -0.18em \PLambda}{}\xspace}

\def\BF         {{\ensuremath{\mathcal{B}}}\xspace}

\def\BR         {\BF}
\def\to                 {\ensuremath{\rightarrow}\xspace}

\def\qsq       {{\ensuremath{q^2}}\xspace}

\def\AT#1     {\ensuremath{A_{\mathrm{T}}^{#1}}\xspace}           % 2

\def\C#1      {\ensuremath{\mathcal{C}_{#1}}\xspace}                       % 9
\def\Cp#1     {\ensuremath{\mathcal{C}_{#1}^{'}}\xspace}                    % 7
\def\Ceff#1   {\ensuremath{\mathcal{C}_{#1}^{\mathrm{(eff)}}}\xspace}        % 9  
\def\Cpeff#1  {\ensuremath{\mathcal{C}_{#1}^{'\mathrm{(eff)}}}\xspace}       % 7
\def\Ope#1    {\ensuremath{\mathcal{O}_{#1}}\xspace}                       % 2
\def\Opep#1   {\ensuremath{\mathcal{O}_{#1}^{'}}\xspace}                    % 7

\newcommand{\tev}{\ifthenelse{\boolean{inbibliography}}{\ensuremath{~T\kern -0.05em eV}}{\ensuremath{\mathrm{\,Te\kern -0.1em V}}}\xspace}
\newcommand{\gev}{\ensuremath{\mathrm{\,Ge\kern -0.1em V}}\xspace}
\newcommand{\mev}{\ensuremath{\mathrm{\,Me\kern -0.1em V}}\xspace}
\newcommand{\kev}{\ensuremath{\mathrm{\,ke\kern -0.1em V}}\xspace}
\newcommand{\ev}{\ensuremath{\mathrm{\,e\kern -0.1em V}}\xspace}
\newcommand{\gevc}{\ensuremath{{\mathrm{\,Ge\kern -0.1em V\!/}c}}\xspace}
\newcommand{\mevc}{\ensuremath{{\mathrm{\,Me\kern -0.1em V\!/}c}}\xspace}
\newcommand{\gevcc}{\ensuremath{{\mathrm{\,Ge\kern -0.1em V\!/}c^2}}\xspace}
\newcommand{\gevgevcccc}{\ensuremath{{\mathrm{\,Ge\kern -0.1em V^2\!/}c^4}}\xspace}
\newcommand{\mevcc}{\ensuremath{{\mathrm{\,Me\kern -0.1em V\!/}c^2}}\xspace}

\def\mum  {\ensuremath{{\,\upmu\mathrm{m}}}\xspace}

\def\gsim{{~\raise.15em\hbox{$>$}\kern-.85em
          \lower.35em\hbox{$\sim$}~}\xspace}
\def\lsim{{~\raise.15em\hbox{$<$}\kern-.85em
          \lower.35em\hbox{$\sim$}~}\xspace}

\def\ptot       {\mbox{$p$}\xspace}

\def\evtgen     {\mbox{\textsc{EvtGen}}\xspace}

\def\photos     {\textsc{Photos}\xspace}

\def\pythia     {\mbox{\textsc{Pythia}}\xspace}

\def\tell1  {TELL1\xspace}
\def\ukl1   {UKL1\xspace}

\newcommand{\eg}{\mbox{\itshape e.g.}\xspace}

\def\RDp      {{\ensuremath{\mathcal{R}(\Dp)}}\xspace}
\def\RDz      {{\ensuremath{\mathcal{R}(\Dz)}}\xspace}
\def\RD       {{\ensuremath{\mathcal{R}(\D)}}\xspace}
\def\Hc       {{\ensuremath{H_c}}\xspace}
\def\Hb       {{\ensuremath{H_b}}\xspace}
\def\RHc       {{\ensuremath{\mathcal{R}(\Hc)}}\xspace}
\def\RDst     {{\ensuremath{\mathcal{R}(\Dstar)}}\xspace}

\def\emax     {{\ensuremath{E_{\rm{max}}}}\xspace}
\def\emu      {{\ensuremath{E_{\mu}}}\xspace}
\def\mmiss    {{\ensuremath{m_{\rm miss}^2}}\xspace}
\def\dQED     {{\ensuremath{\delta_{\rm{QED}}}}\xspace}
\def\OmegaC   {{\ensuremath{\Omega_{\rm{C}}}}\xspace}
\def\qsq      {{\ensuremath{q^2}}\xspace}

\usepackage{mciteplus}
\newboolean{inbibliography}
\setboolean{inbibliography}{false} %True once you enter the bibliography
\usepackage{cite} % Allows for ranges in citations

\journalname{Eur. Phys. J. C}

\begin{document}

\title{Impacts of radiative corrections on measurements of lepton flavour universality in $\B \to \D \ell \neul$ decays}

\author{Stefano Cal\'{i}\thanksref{e1,addr1}
        \and
        Suzanne Klaver\thanksref{e2,addr1}
        \and
        Marcello Rotondo\thanksref{e3,addr1}
        \and
        Barbara Sciascia\thanksref{e4,addr1}
}

\thankstext{e1}{e-mail: stefano.cali@cern.ch}
\thankstext{e2}{e-mail: suzanne.klaver@cern.ch}
\thankstext{e3}{e-mail: marcello.rotondo@cern.ch}
\thankstext{e4}{e-mail: barbara.sciascia@cern.ch}

\institute{INFN Laboratori Nazionali di Frascati, Via Enrico Fermi, 40, 00044 Frascati, Italy\label{addr1}
}

\date{Received: 28 May 2019 / Accepted: 27 August 2019}

\maketitle

\begin{abstract}

Radiative corrections to $\B \to \D \ell \neul$ decays may have an impact on predictions and measurements of the lepton flavour universality observables \RDp and \RDz.
In this paper, a comparison between recent calculations of the effect of soft-photon corrections on \RDp and \RDz, and corrections generated by the widely used package \photos is given. 
The impact of long-distance Coulomb interactions, which are not simulated in \photos, is discussed.  
Furthermore, the effect of high-energy photon emission 
is studied through pseudo-experiments in an \lhcb-like environment. It is found that 
over- or underestimating these emissions can cause a bias on \RD  as high as 7\%. 
However, this bias depends on individual analyses, and future high precision measurements  require an accurate evaluation of these QED corrections.

\end{abstract}

%-----------------------------------------------------------------------------------

\section{Introduction}
The Standard Model (SM) assumes lepton universality (LU) implying that once the mass difference is taken into account, all SM interactions treat the three charged leptons identically. The mass difference results in a different phase space between decays involving \taum and the lighter \en and \mun leptons\footnote{Throughout this paper, the inclusion of charge-conjugate processes is implied and natural units with $\hbar=c=1$ are used.}. LU can be tested by measuring the ratio of decay rates, ensuring that the Cabibbo-Kobayashi-Maskawa matrix elements, as well as most of the form factors, cancel in the ratio. This results in more accurate theoretical predictions and in the cancellation of many experimental systematic uncertainties. 
One type of these LU measurements is performed using semileptonic \B decays of the form
$\bquark\to\cquark\ellm\neulb$, commonly known as measurements of \RHc, defined as
\begin{align}
    \RHc &= \frac{\BR(\Hb\to\Hc\taum\neutb)}{\BR(\Hb\to\Hc\ell^-\neulb)} \, ,
\end{align}
where \Hb and \Hc are a \bquark and \cquark hadron, respectively, and $\ell$ is either 
an electron or muon.

Several measurements of \RHc have been performed by the \lhcb, \belle and \babar experiments. 
For \RD, on which this paper is focused, the predicted value~\cite{Bigi:2016mdz,Bernlochner:2017jka,Jaiswal:2017rve,Aoki:2019cca} is 
\begin{align}
\RD = \RDp = \RDz = 0.299 \pm 0.003 \, ,
\end{align}
which assumes isospin symmetry. The average of the measured value of \RD is
$0.349\pm 0.027\pm 0.015$ \cite{Lees:2012xj,Huschle:2015rga,Abdesselam:2019dgh}, 
where the first uncertainty is statistical and the second systematic. 
Even though \RD differs from the SM prediction by only 1.4$\sigma$, it is remarkable that the deviation from the SM of the combined \RD and \RDst observables is 3.1$\sigma$~\cite{HFLAV16}.

Radiative corrections were long thought to be negligible
at the level of precision of measurements and predictions of \RD.
Recently, however, de Boer et al.~\cite{deBoer:2018ipi} presented a new evaluation of the 
long-distance electromagnetic (QED) contributions
to $\Bzb\to\Dp\ellm\neulb$ and $\Bm\to\Dz\ellm\neulb$ decays, where $\ellm = \mun, \taum$. 
They point out that these soft-photon corrections are different for \mun and \taum decays, such that they do not cancel in the ratios \RDp and \RDz. According to the authors of Ref.~\cite{deBoer:2018ipi}, the proper evaluation of the radiative corrections alters the SM predictions of the \RD and \RDst values and increases their uncertainty.
The current tension between the SM and measurements could be weakened or strengthened if radiative corrections are not properly taken into account. 

All experiments measuring these types of LU are dependent on the simulation of 
QED radiative corrections in decays of particles and resonances. 
The widely used package to simulate these corrections is  \photos~\cite{Barberio:1993qi,Golonka:2005pn}, which is
used by all three experiments measuring \RD and \RDst. 

This paper starts by comparing the radiative corrections on \RDp and \RDz from Ref.~\cite{deBoer:2018ipi} with those simulated by \photos in Sect.~\ref{sec:comparePhotos}. 
The sensitivity of measurements of \RDp and \RDz to radiative corrections in the \mun and \taum decay modes is studied with pseudo-experiments in an \lhcb-like environment, with different assumptions on the shape of the total energy of the radiated photons.
The method and the results of this study are reported in Sect.~\ref{sec:dummyAnalysis}. Conclusions and recommendations are summarised in Sect.~\ref{sec:Conclusions}.

%-----------------------------------------------------------------------------------

\section{Radiative corrections in P{\small HOTOS}}
\label{sec:comparePhotos}

\photos~\cite{Golonka:2005pn,Golonka:1379813} is a universal Monte Carlo algorithm that simulates the effects of QED corrections in decays of particles and resonances. It exploits the factorisation property of QED coming from the exponentiation method used to improve the convergence of the perturbative expansion. Any particle-decay process accompanied by bremsstrahlung photons can be factorised into a tree term and bremsstrahlung factor. The latter depends only on the four-momenta of those particles taking part in the decay, and not on the underlying process. This approximation, which takes into account both real and virtual corrections, converges to an exact expression in the soft-photon region of phase space. 
It is worth noting that \photos does not incorporate the emission of photons depending on the hadronic structure. These so called structure-dependent (SD) photons impact the spin of the decay particle, and may also interfere with bremsstrahlung photons. The effect of SD photons depends on the specific decay under study and, as was the case for kaon decays~\cite{Bijnens:1994me}, may not be negligible.

The latest versions of \photos include multi-photon emissions, and interference between final-state photons.
The validity of \photos has been tested successfully by comparing its results to full calculations available in various processes involving \W, \Z and hadronic \B decays into scalar mesons~\cite{Golonka:2005pn,Nanava:2006vv}. 
Because of the universal treatment of photon emission in \photos, its performances in specific processes should always be checked, especially when high precision is desired or when signal extraction is sensitive to detailed simulation of a phase space corner of the decay.

The calculation by de Boer et al. in Ref.~\cite{deBoer:2018ipi} is the first that studies the impact of soft-photon corrections on \RDp and \RDz. It is valid in the regime
in which the maximum energy of the radiated photons is smaller than the lepton mass,
which is the muon mass in this case.
This calculation includes more effects than \photos does, in particular the interference between initial- and final-state photons, and the Coulomb correction. The latter increases the decay rate of decays with charged particles in the final state. It should be noted that the contribution of the Coulomb correction is singular for null relative velocity between final-state charged particles.

To compare QED corrections between \photos and Ref.~\cite{deBoer:2018ipi}, four samples ($\Bzb\to\Dp\ellm\neulb$ and $\Bm\to\Dz\ellm\neulb$, where $\ellm = \mun, \taum$) with three million \B-meson decays are generated by \pythia 8~\cite{Sjostrand:2006za, Sjostrand:2007gs}. The decays are simulated by \evtgen~\cite{Lange:2001uf}, and the radiative corrections by \photos v.3.56, with the ``option with interference'' switched on. QED corrections are applied by \photos by modifying the charged track's four-momentum in the event record filled by \evtgen every time a photon is added.

The four-momentum of the total radiated photons, $\ptot_{\gamma}$, is defined as 
\begin{equation}
    \ptot_{\gamma} = \ptot_{B} - \left(\ptot_{\D} + \ptot_{\ellm} + \ptot_{\neulb} \right) \, ,
\end{equation}
where $\ptot_{B}$, $\ptot_{\D}$, $\ptot_{\neulb}$, and $\ptot_{\ellm}$ are the four-momenta of the 
\B, \D, \ellm and \neulb particles, respectively, taken from the event record updated by \photos. 
This means that, in agreement with 
Ref.~\cite{deBoer:2018ipi}, the radiation of the \D decay products is not taken into account. 
The total energy of the radiated photons, $E_{\gamma}$, is computed in the \B rest frame.
As in Ref.~\cite{deBoer:2018ipi}, the variable \emax is defined as the maximum value that $E_{\gamma}$ is allowed to have to consider $\B\to\D\ell{\bar{\nu}}_\ell(\gamma)$ as signal.

\begin{figure}[tb]
  \begin{center}
    \includegraphics[width=\linewidth]{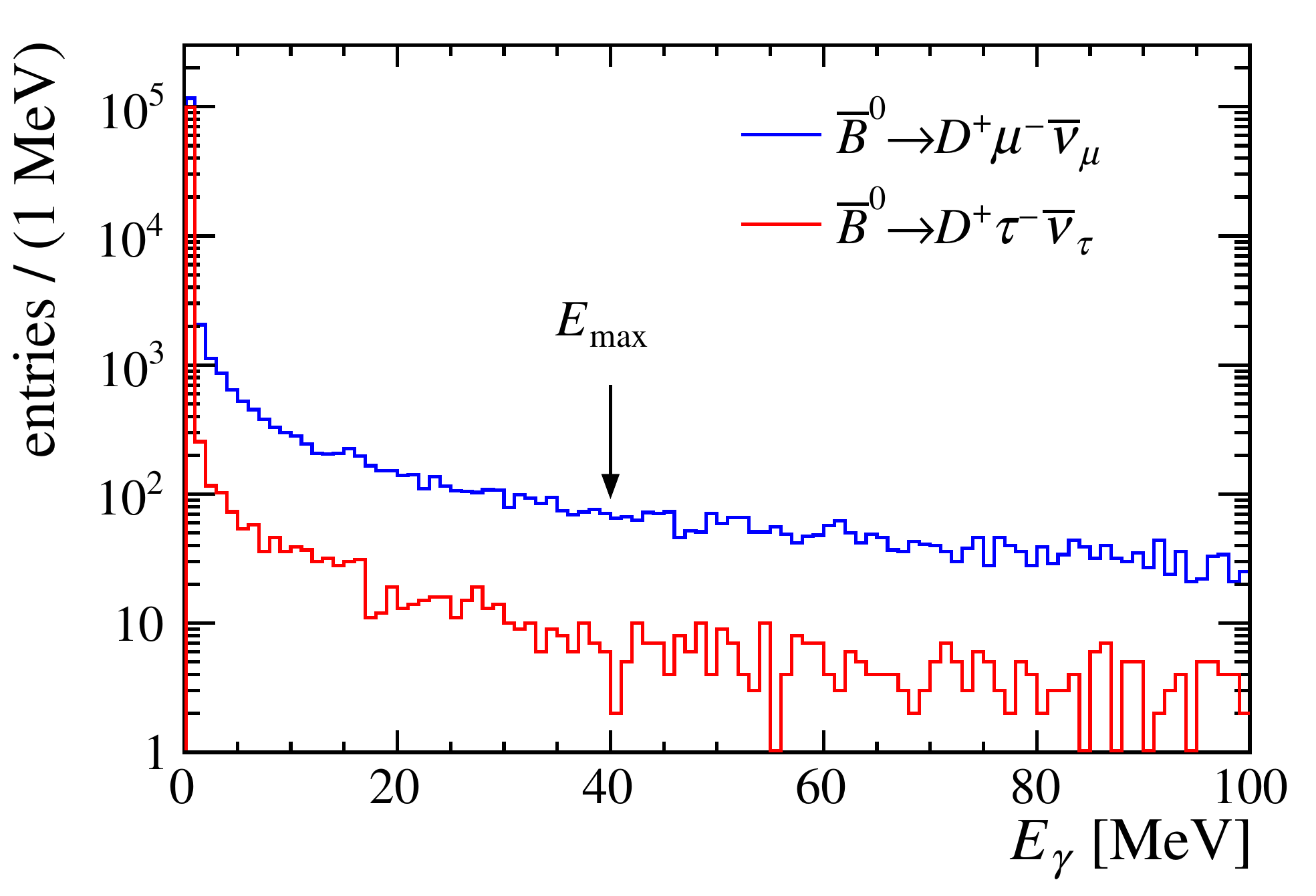} 
  \end{center}
  \caption{
Distribution of the total energy of the radiated photons, $E_{\gamma}$, up to 100~MeV for \Bzb\to\Dp\taum\neutb and \Bzb\to\Dp\mun\neumb decays as simulated by \photos. Once a value of \emax is chosen, \eg 40~MeV as in the plot, all events with higher $E_{\gamma}$ values are discarded.
}
  \label{fig:Emax}
\end{figure}

The QED correction, \dQED, is given by the relative variation of the branching ratio when events with total radiated energy greater than \emax are discarded. This can be calculated as follows:
\begin{equation}
\dQED = \frac{\int_{0}^{\emax} N(E_{\gamma} ) dE_{\gamma} }{\int_{0}^{\infty} N(E_{\gamma}) dE_{\gamma}} - 1\, ,
\end{equation}
where $N(E_{\gamma} )$ is the distribution of events with $E_{\gamma}$. This distribution is shown for \Bzb\to\Dp\taum\neutb and \Bzb\to\Dp\mun\neumb decays in Fig.~\ref{fig:Emax}.
The considered energy range is up to 100~MeV, which covers the majority of radiative photons, namely 98\% of the \mun decays and 99.7\% for the \taum decays generated by \photos.

\begin{figure*}[tb]
  \begin{center}
    \includegraphics[width=0.32\linewidth]{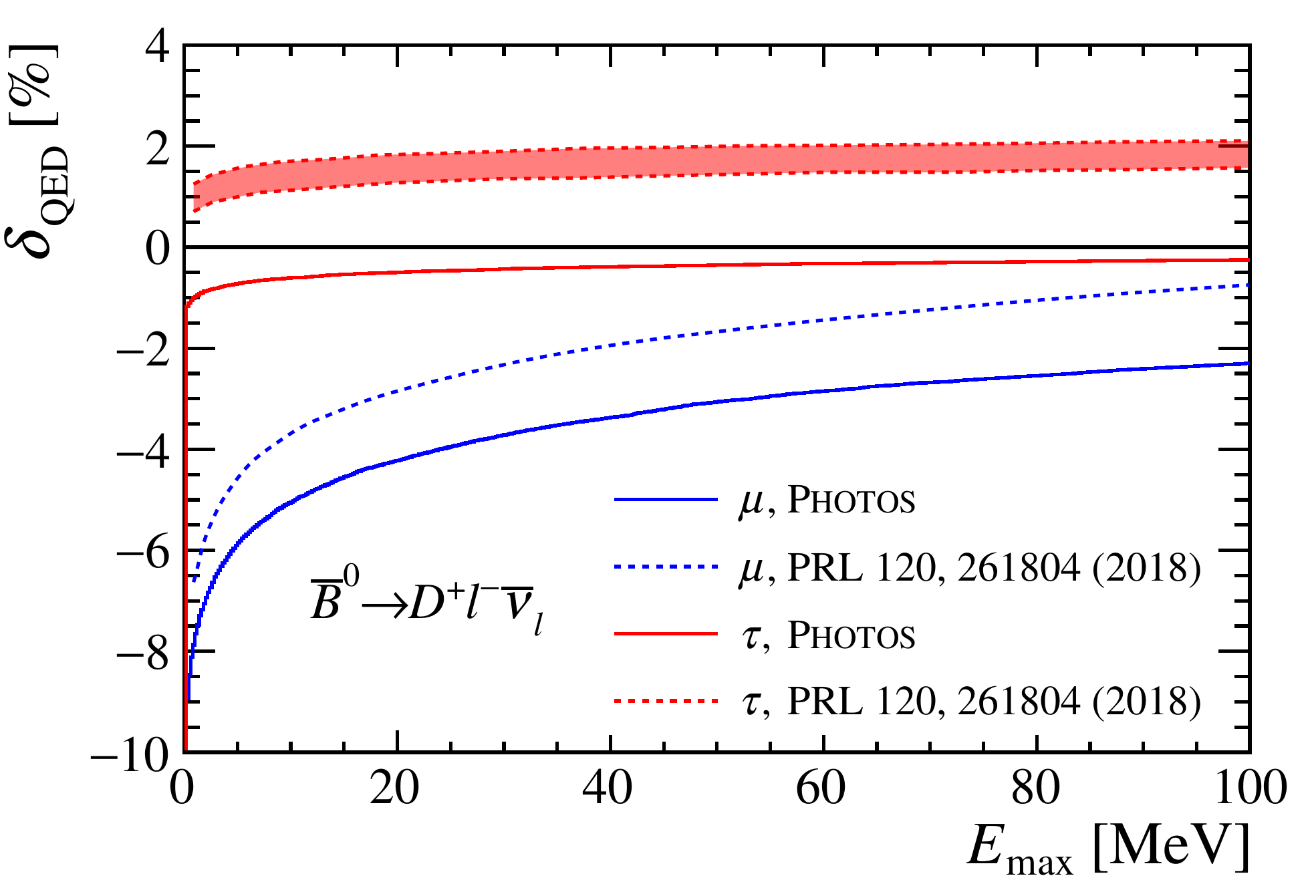} 
    \includegraphics[width=0.32\linewidth]{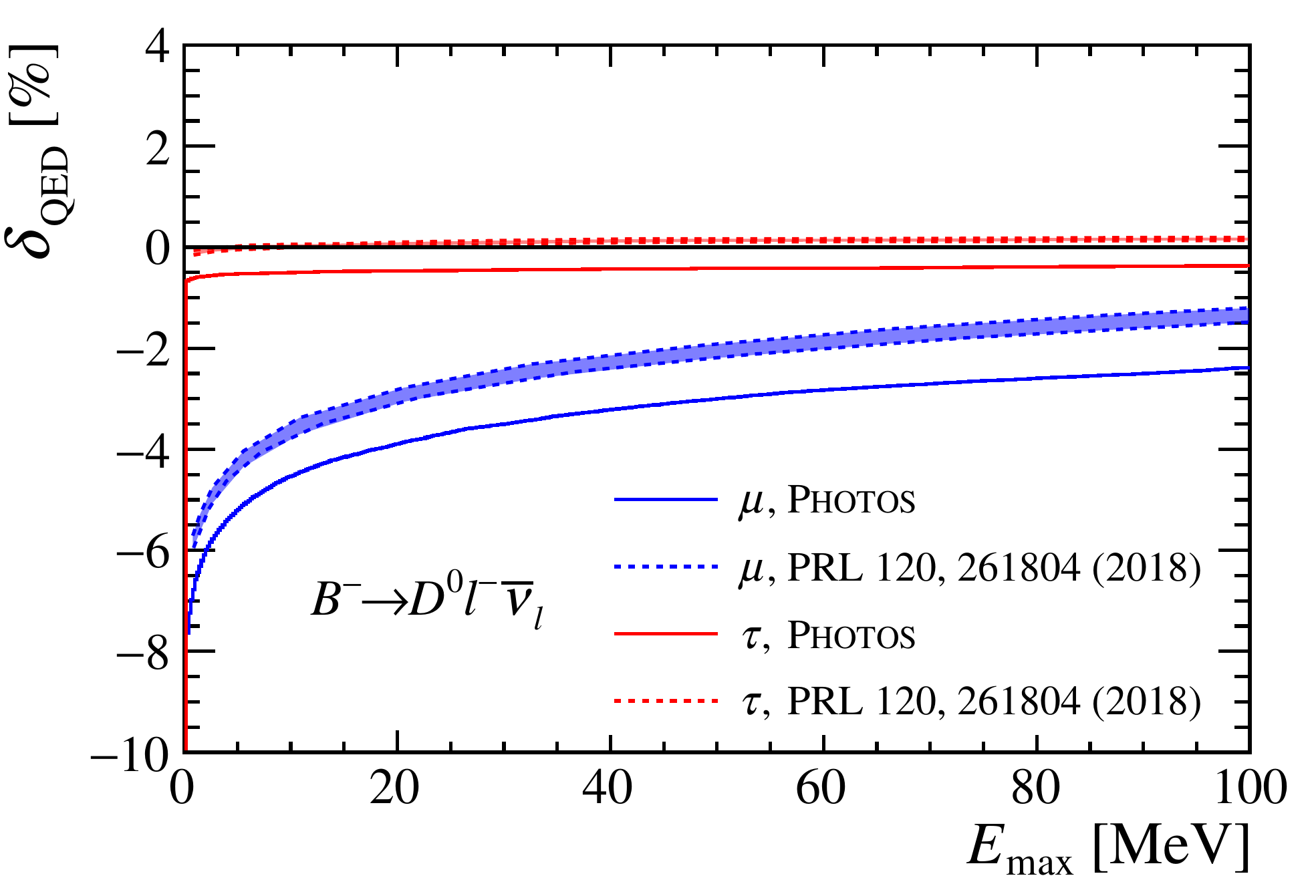} 
    \includegraphics[width=0.32\linewidth]{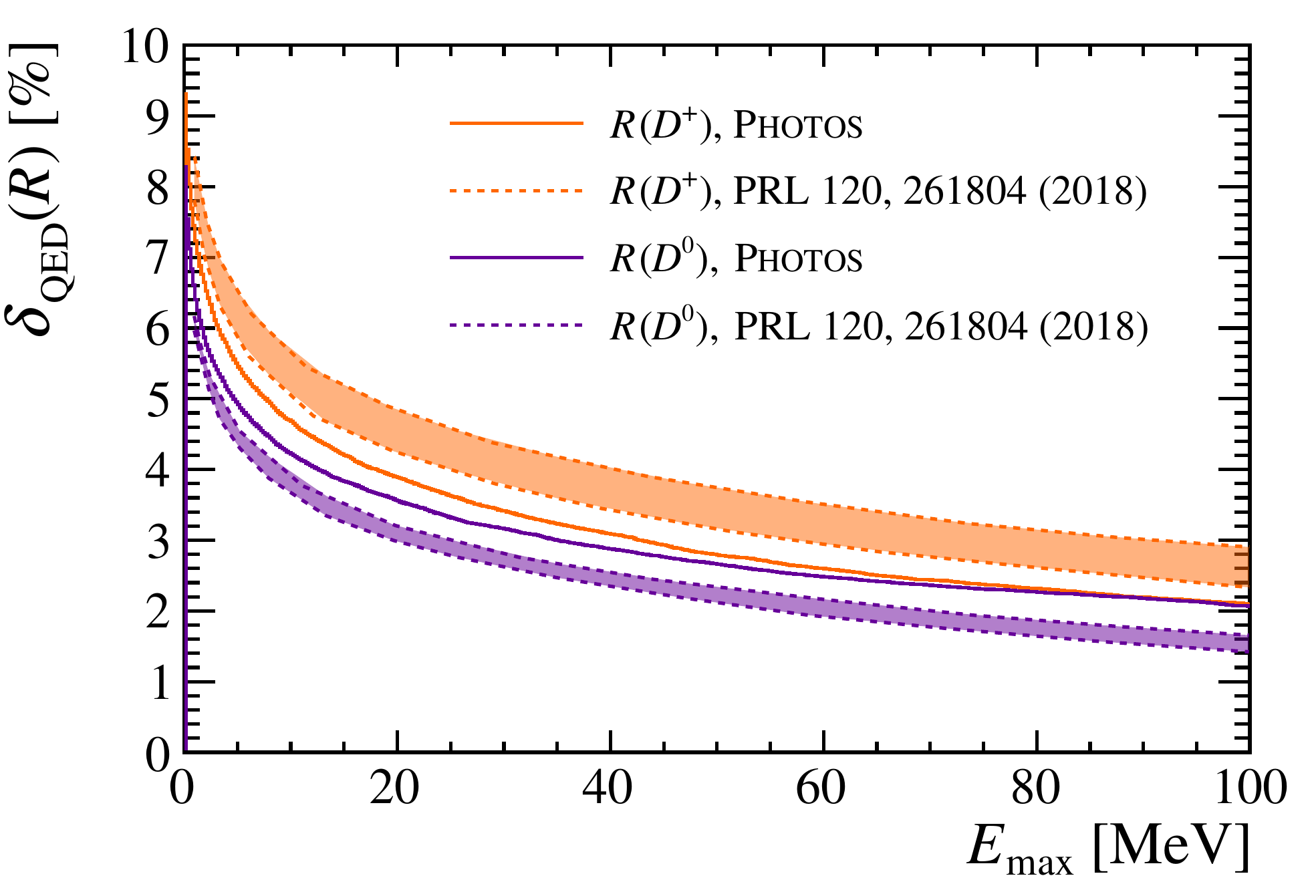}
  \end{center}
  \caption{
  Radiative corrections to the branching ratios of 
  $\bar{B}^0\to D^+\ell^-\bar{\nu_\ell}$ (left) and $B^-\to D^0\ell^-\bar{\nu_\ell}$ (middle) decays, as a function of \emax. 
  The long-distance QED corrections to \RDp (orange) and \RDz 
  (violet) as a function of \emax (right).
  The plots are obtained from simulated data (solid 
  lines, which include the statistical uncertainty) and from Ref.~\cite{deBoer:2018ipi} (dashed lines, filled with transparent 
  colours when the uncertainties are significant).}
  \label{fig:RD0plots}
\end{figure*}

Comparisons between radiative corrections from \photos 
and Ref.~\cite{deBoer:2018ipi} are shown in Fig.~\ref{fig:RD0plots} for the 
$\Bzb\to\Dp\ellm\neulb$ (left panel) and $\Bm\to\Dz\ellm\neulb$ (middle panel) branching fractions. 
These plots show differences of up to 2\% for \Bzb decays, and $0.5-1\%$ for \Bm decays. 
Unfortunately, the effect does not cancel in the ratios of branching fractions. This is clearly visible from the right panel of Fig.~\ref{fig:RD0plots} where radiative corrections on \RD, $\delta_{QED}(\mathcal{R})$, are shown as a function of \emax. 
\photos predicts a QED correction that is 0.5\% lower than the one in Ref.~\cite{deBoer:2018ipi} for \RDp, while it is 0.5\% higher than the one in Ref.~\cite{deBoer:2018ipi} for \RDz.

\subsection{Coulomb correction}
A significant part of the radiative corrections in Ref.~\cite{deBoer:2018ipi} originates
from Coulomb interactions, which are not included in \photos. Note that the
Coulomb correction is relevant for the \Dp mode, but not for the \Dz mode.
For a fermion-scalar (and fermion-fermion) pair, this correction is given by
\begin{equation}
   \OmegaC = \frac{2\pi\alpha}{\beta_{D\ell}} \frac{1}{1-e^{-\frac{2\pi\alpha}{\beta_{D\ell}}}}{\mbox ,} 
\end{equation}
where $\alpha = 1/137$ and $\beta_{D\ell}$ is the relative velocity between 
the \D meson and the lepton, defined as
\begin{equation}
    \beta_{D\ell} = \left [ 1 - \frac{4m_D^2 m_\ell^2}{(s_{D\ell}-m_D^2 - m_\ell^2)^2} \right ]^{1/2} \, ,
\end{equation}
where $s_{D\ell} = (p_D+p_\ell)^2$. A well-known approximation of the Coulomb correction by Atwood and Marciano~\cite{Atwood:1989em}, yields $\OmegaC = (1+\pi\alpha) \, \approx 1.023$ 
which occurs when $\beta_{D\ell} \approx 1$. This is accurate for decays with light leptons, 
but not for those with \taum leptons.  
For the semitauonic mode, the typical relative velocity is 0.5-0.9, 
resulting in a Coulomb correction between 2.5 and 5.0\%.

\begin{figure*}[tb]
  \begin{center}
    \includegraphics[width=0.32\linewidth]{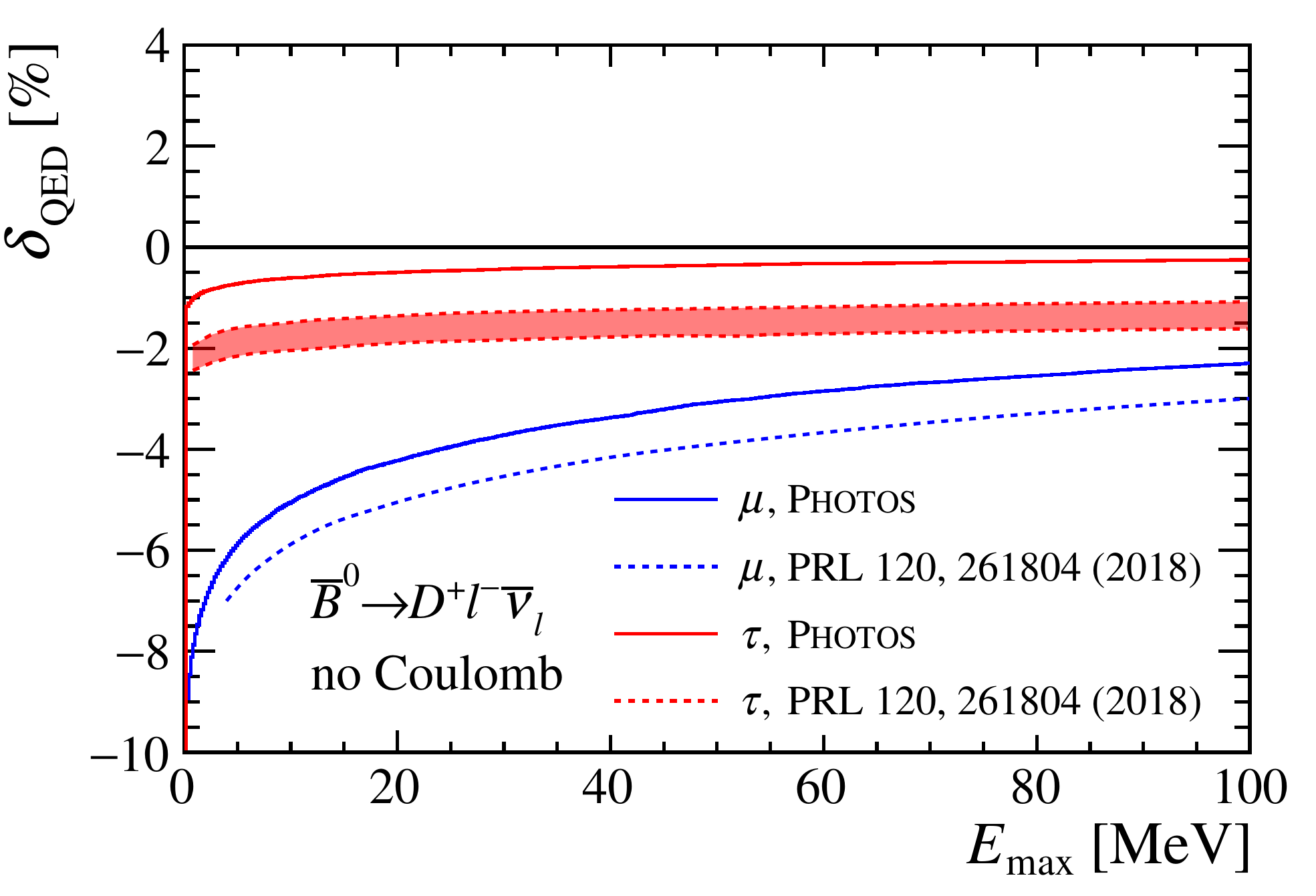}
    \includegraphics[width=0.32\linewidth]{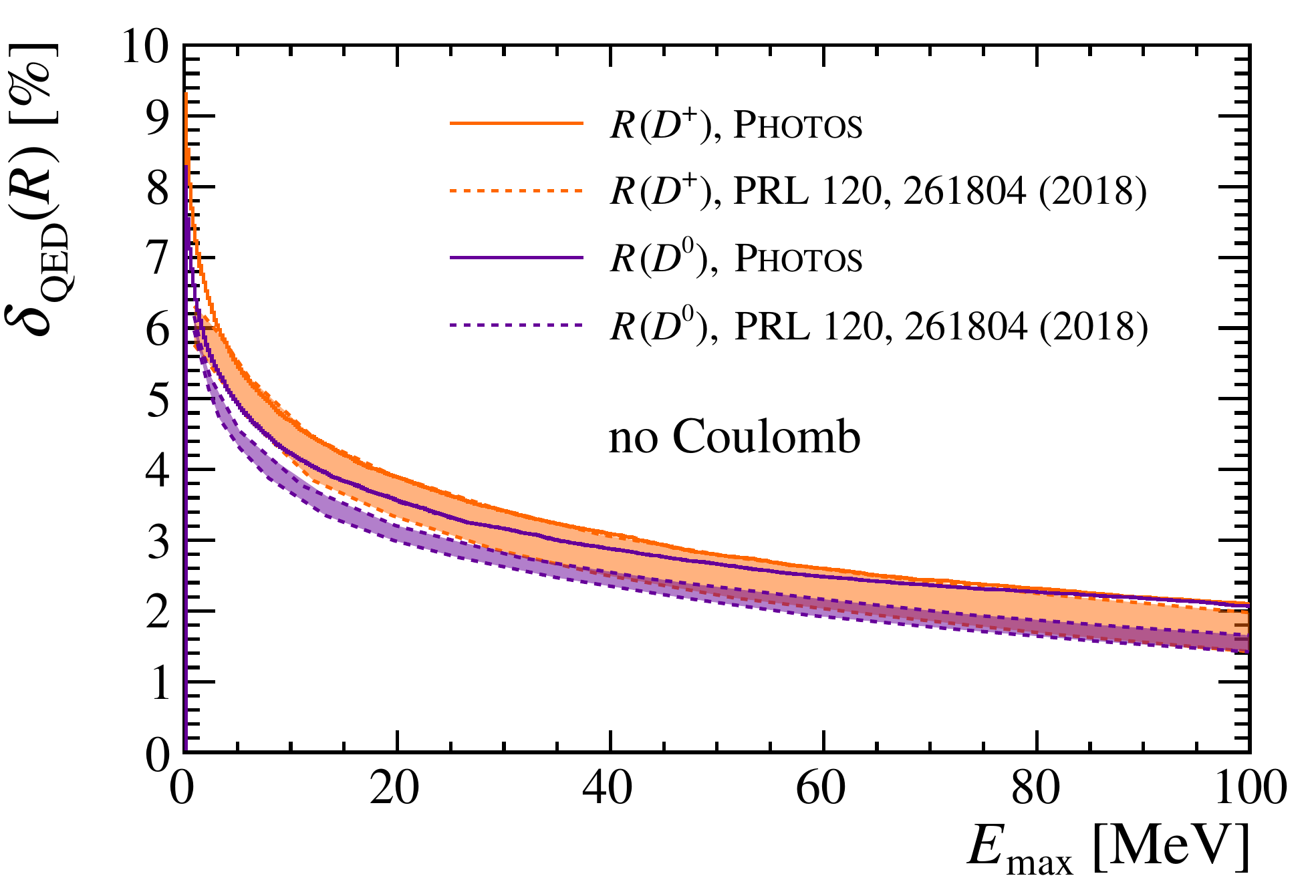}
    \includegraphics[width=0.32\linewidth]{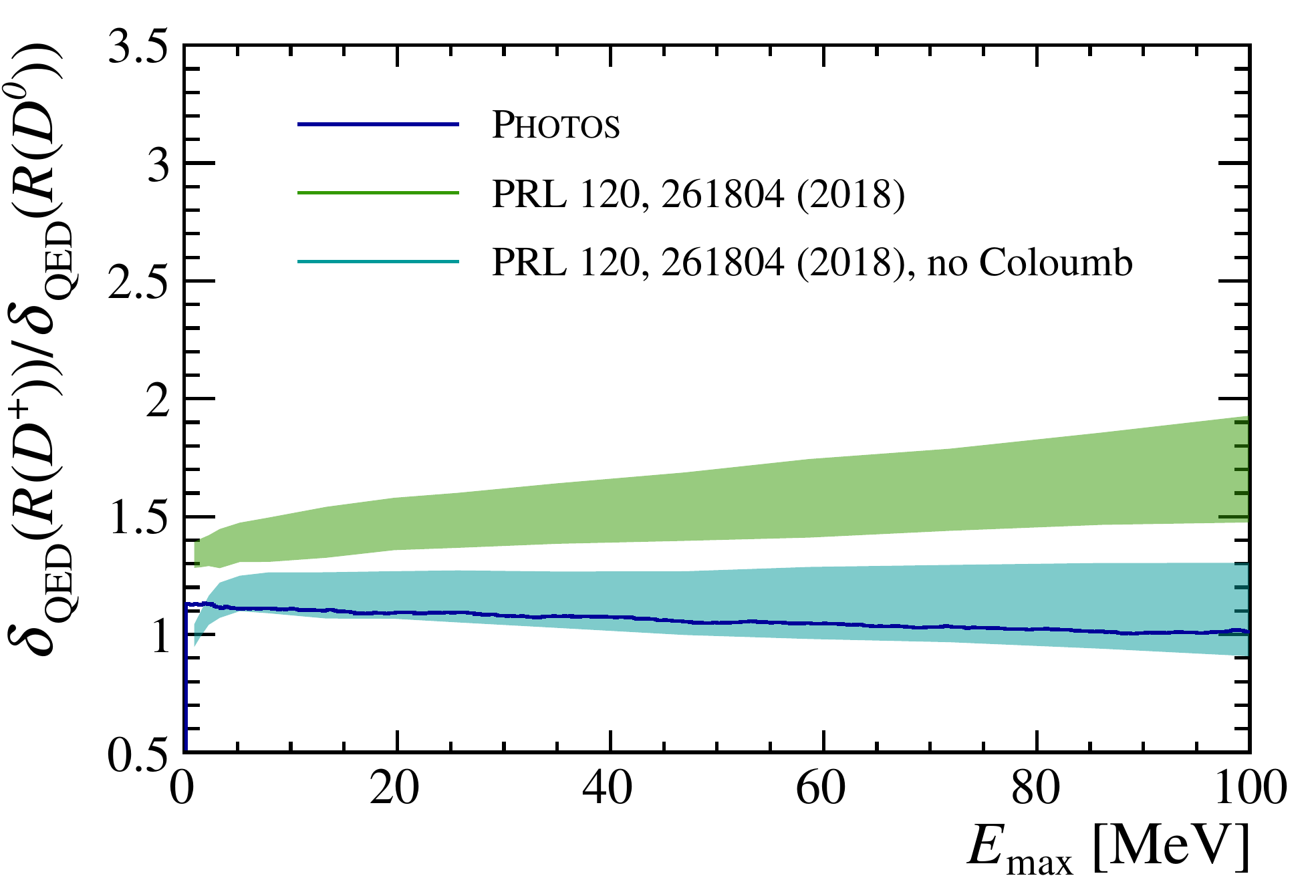}
  \end{center}
  \caption{
  Radiative corrections to the branching ratios of 
  $\bar{B}^0\to D^+\ell^-\bar{\nu_\ell}$ (left) and \RDp (middle)
  in the case that no Coulomb correction is applied. The plot on the right shows the ratio 
  $\dQED(\RDp)/\dQED(\RDz)$. The plots are 
  obtained from simulated data (solid lines, which include the statistical uncertainty) and from 
  Ref.~\cite{deBoer:2018ipi} (dashed lines, filled with transparent colours when the 
  uncertainties are significant). }
  \label{fig:noCoulomb}
\end{figure*}

QED corrections from \photos for the \Dp mode 
are also compared with predictions not including the Coulomb correction from Ref.~\cite{deBoer:2018ipi}.
This reduces the difference of the corrections to the branching ratios between \photos and the theoretical calculations to about 1\% and brings the corrections on \RDp in close agreement, as shown in Fig.~\ref{fig:noCoulomb} (left and middle, respectively).

Fig.~\ref{fig:noCoulomb} (right) shows the ratio of QED corrections on \RDp over those on \RDz. It is worth noting that both \photos and the calculation in Ref.~\cite{deBoer:2018ipi} without Coulomb correction conserve isospin symmetry (\dQED values for \RDp and \RDz agree within the errors), while the Coulomb correction introduces an isospin-breaking term.

%-----------------------------------------------------------------------------------

\section{Effects on LHCb-like analysis}
\label{sec:dummyAnalysis}

The comparison in the previous section only holds for values of $E_{\gamma}$ up to 100~MeV. For higher energies, no calculations relevant to \RD are available\footnote{In Ref.\cite{Bernlochner:2010yd} a calculation of the high-energy SD contribution to $\B\to\D\ell\neulb$ is reported. This is not relevant to this study because of the missing lepton-mass dependent effects.}. Nevertheless, \photos generates photons with energies larger than those treated in Ref.~\cite{deBoer:2018ipi}, which is the range where the effect of SD photons might be relevant.

To study the effect of under- or overestimating radiative corrections in simulations used for measurements of \RD, a simplified analysis is performed in an \lhcb-like environment. Also here, the radiation emitted by the decay products of the \D mesons is neglected because their contributions largely cancel out in the ratio. 

The strategy of this study consists of fitting a data sample with templates describing the $\B\to\D\mun\neumb$ and $\B\to\D\taum\neutb$ components. The fits are performed with templates built under the hypothesis that no radiation with $E_\gamma$ above a certain value \emax is emitted. 
In particular, five \emax values were chosen to cut on $E_\gamma$: 100, 300, 500, 800 and 1500~MeV. The bias on \RD, determined from these fits, is an indication of the importance of the simulation of the $E_\gamma$ distribution in the high-energy region.

This analysis follows a strategy similar to the one used in Ref.~\cite{Aaij:2015yra}, where \RDst is measured using a three-dimensional templated fit. The data samples, referred to as pseudo-experiments in the following text, are generated from a mixture of $\B\to\D\mun\neumb$ and $\B\to\D\taum\neutb$ decays, with radiative corrections generated by \photos. Here \RD is assumed to be 0.3 as predicted by the SM. 

The variables used in the templated fit performed to extract the $\B\to\D\mun\neumb$ and $\B\to\D\taum\neutb$ yields from the pseudo-experiments are: the muon energy computed in the \B meson rest frame, \emu; the missing mass squared, $\mmiss = (\ptot_{\B}-\ptot_{\D}-\ptot_{\mu})^2$; and the squared four-momentum transferred to the lepton system, $\qsq=(\ptot_{\B}-\ptot_{\D})^2$. The
variables are binned as follows: four bins in \qsq in the range $-0.4<\qsq<12.6$ GeV, 40 bins in \mmiss between $-2<\mmiss<10$ GeV$^2$, and 30 bins in muon energy in the range $100<\emu<2500$ MeV, consistent with Ref.~\cite{Aaij:2015yra}. In this case study, only the signal ($\B\to\D\taum\neutb$) and normalisation ($\B\to\D\mun\neumb$) components are considered, while all backgrounds are ignored.

Basic selection requirements are applied to mimic the acceptance of the \lhcb detector and its trigger following Ref.~\cite{Ciezarek:2016lqu}. 
Both production and \B-decay vertex positions are smeared to simulate the resolution of the \lhcb
detector. The resolution on the production vertices
is 13\mum in $x$ and $y$, and 70\mum in $z$ direction. For the \B-decay vertices a
resolution of 20\mum in $x$ and $y$, and 200\mum in $z$ direction is used, after which the \B direction is computed. 
The \D mesons decay as $\Dz\to\Km\pip$ and $\Dp\to\Km\pip\pip$, the \taum lepton as
$\taum\to\mun\neumb\neut$. The muons and all decay products from the \D mesons are
required to be in the pseudorapidity range between $1.9$ and $4.9$.
In addition, the momentum of each of these particles is required to be 
larger than $5$ GeV, and its component transverse to the beam direction must be larger than $250$ MeV.
The distance between the production and \B-decay vertex should be at least 3 mm, similar to
the requirements applied in a typical trigger selection.

\begin{figure*}[tb]
  \begin{center}
    \includegraphics[width=0.32\linewidth]{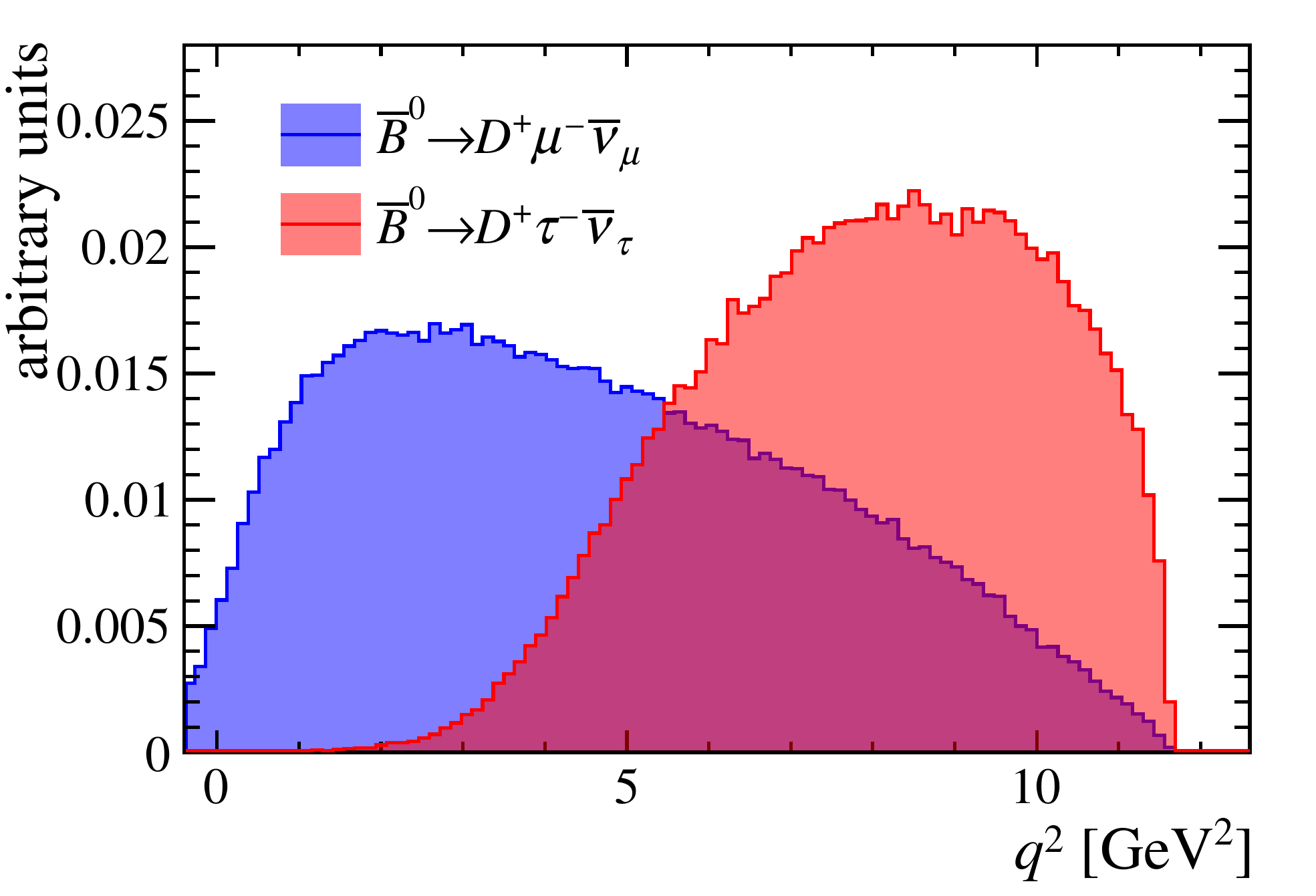} 
    \includegraphics[width=0.32\linewidth]{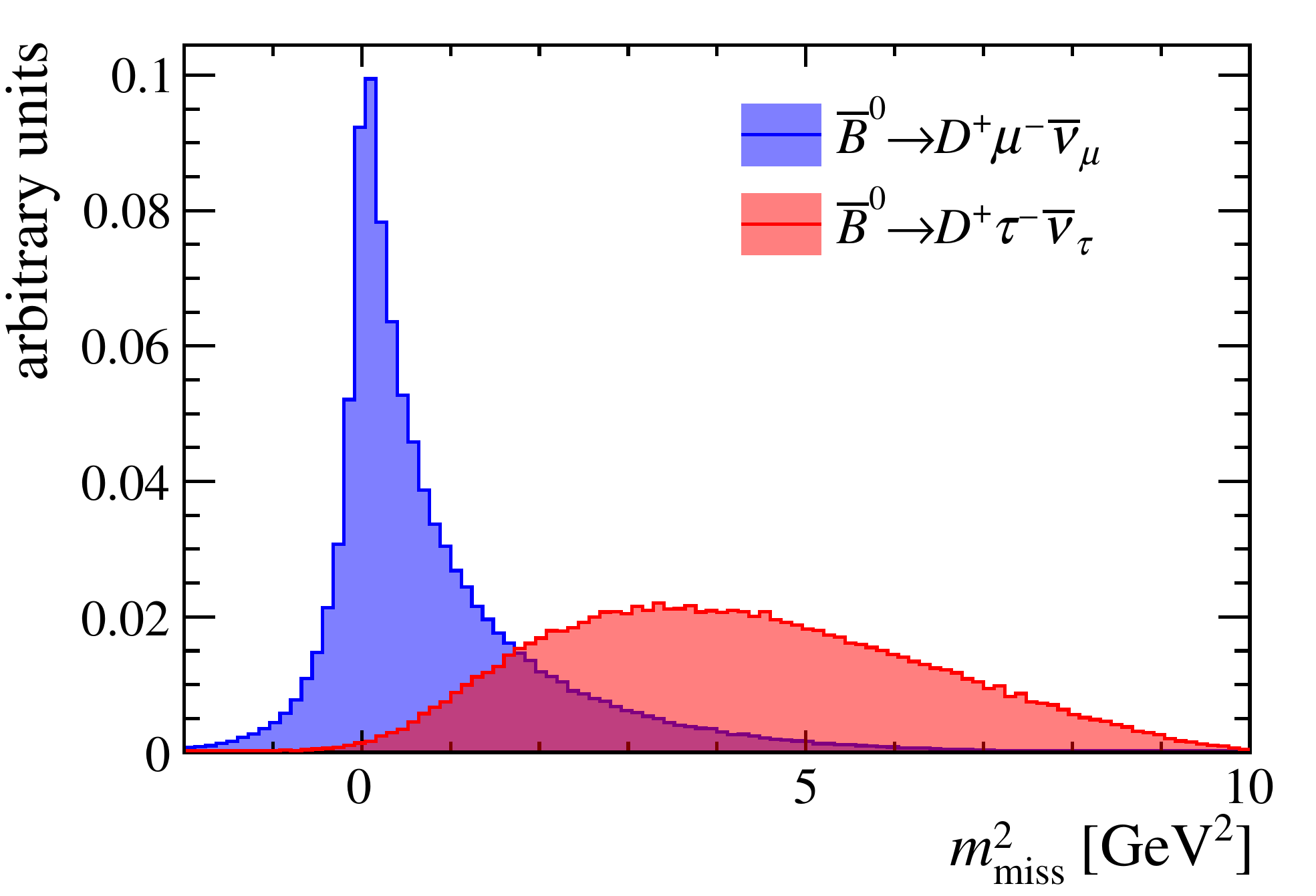}
    \includegraphics[width=0.32\linewidth]{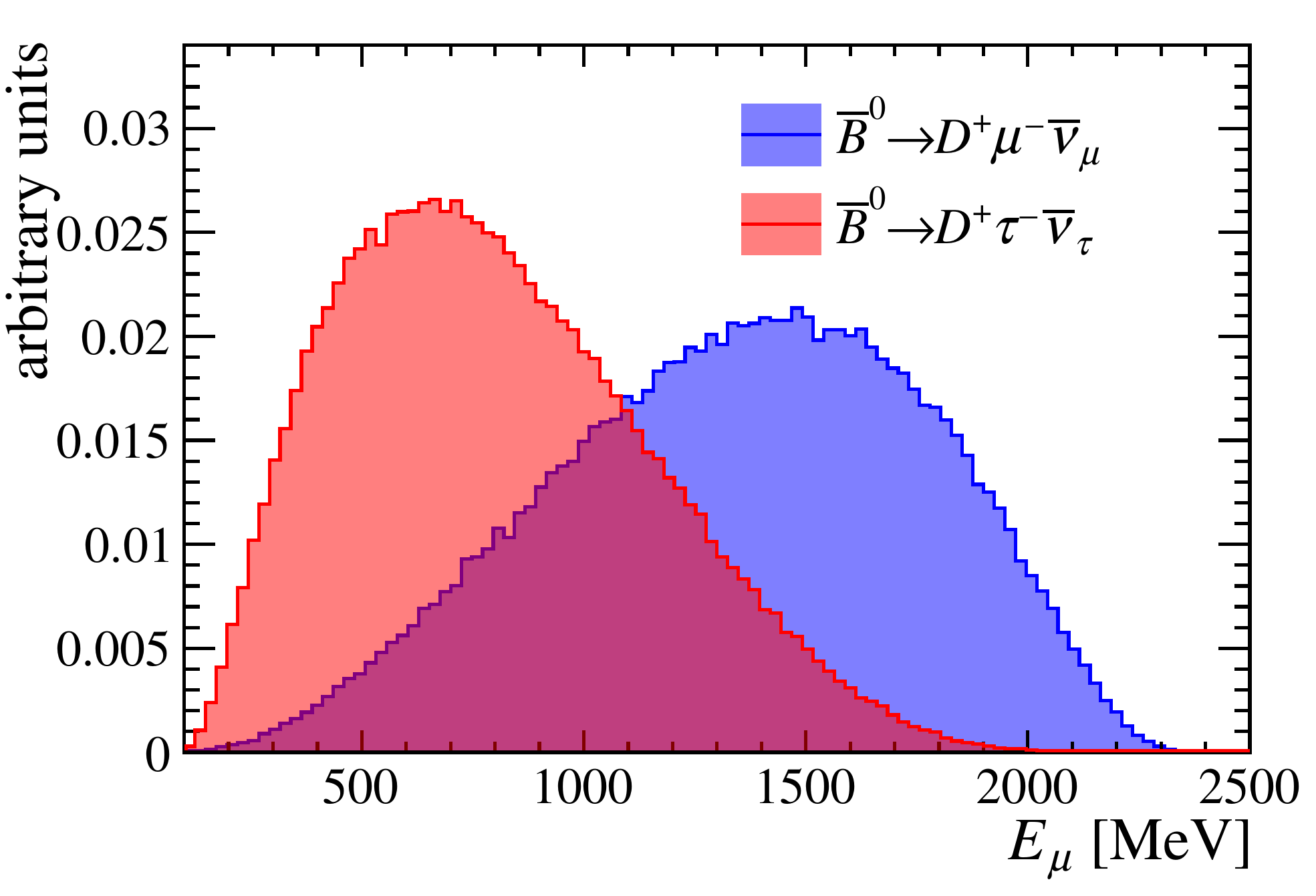} 
  \end{center}
  \caption{Shapes of the \qsq, \mmiss, and \emu distributions after applying basic selection requirements and the
  rest frame approximation for the $\Bzb\to\Dp\ellm\neulb$
  decays where $\ellm = \mun, \taum$.}
  \label{fig:RF_shapes}
\end{figure*}

Due to the missing neutrino and unknown effective centre-of-mass energy of the collision, the \B-meson momentum cannot be reconstructed in an \RHc analysis at \lhcb.
Therefore the momentum of the \B meson in the $z$ direction, $(\ptot_{\B})_z$, is approximated as 
$(\ptot_{\B})_z=(m_{\B}/m_{\rm vis})(\ptot_{\rm vis})_z$, where $m_{\B}$ is the \B mass,
and $m_{\rm vis}$ and $(\ptot_{\rm vis})_z$ are the momentum in the $z$ direction and the mass of the
visible decay products of the \B meson, respectively. This directly follows the
approach from Ref.~\cite{Aaij:2015yra}. After computing the \B momentum with the above approximation and applying 
the selection criteria described in this section, \qsq, \mmiss and \emu 
are calculated. The distributions for the signal and control samples are shown in Fig.~\ref{fig:RF_shapes}.
Even using this simplified detector description, these distributions show the same key features as those in Ref.~\cite{Aaij:2015yra}.

\begin{figure*}[tb]
  \begin{center}
    \includegraphics[width=0.48\linewidth]{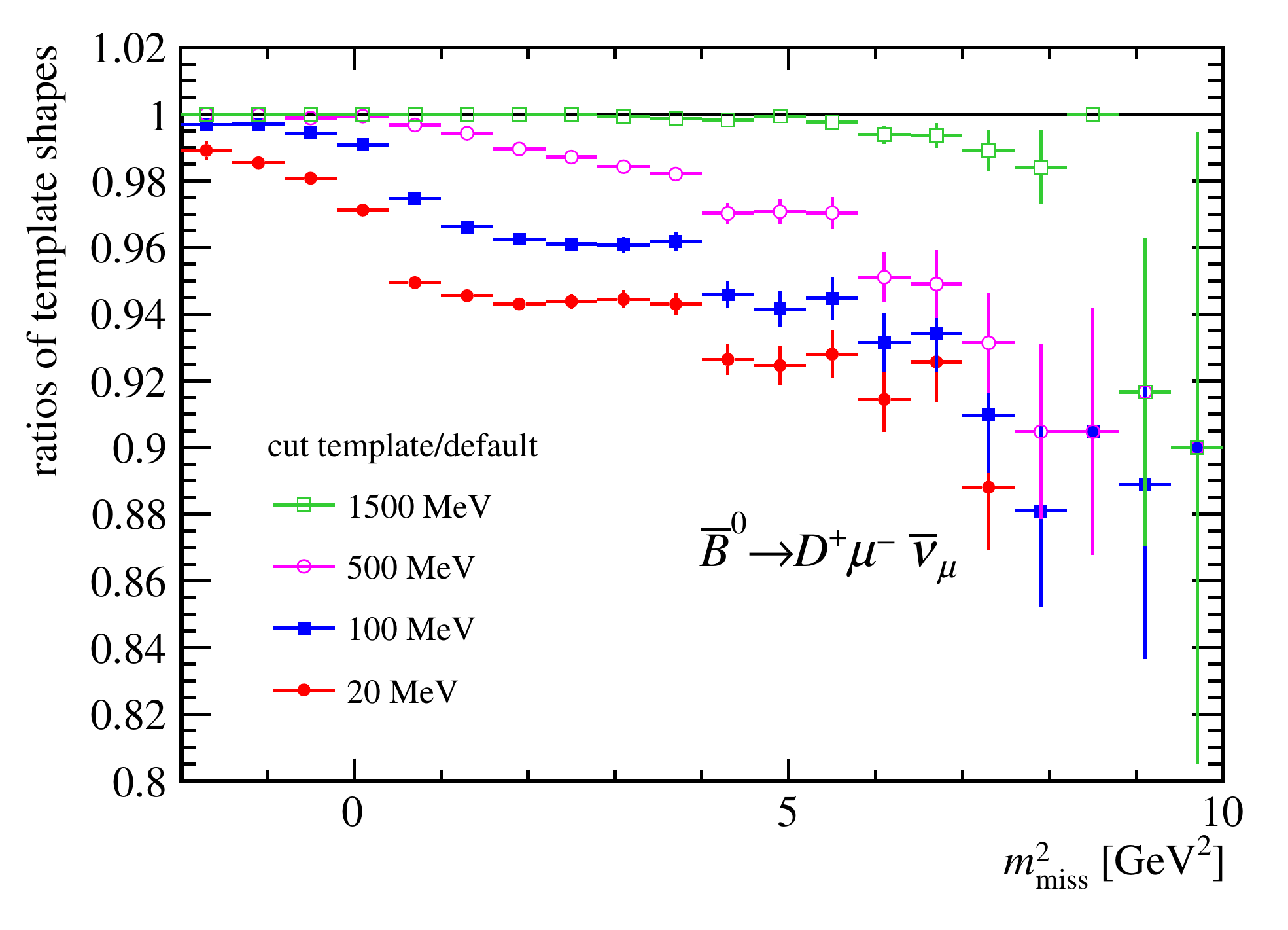}
    \includegraphics[width=0.48\linewidth]{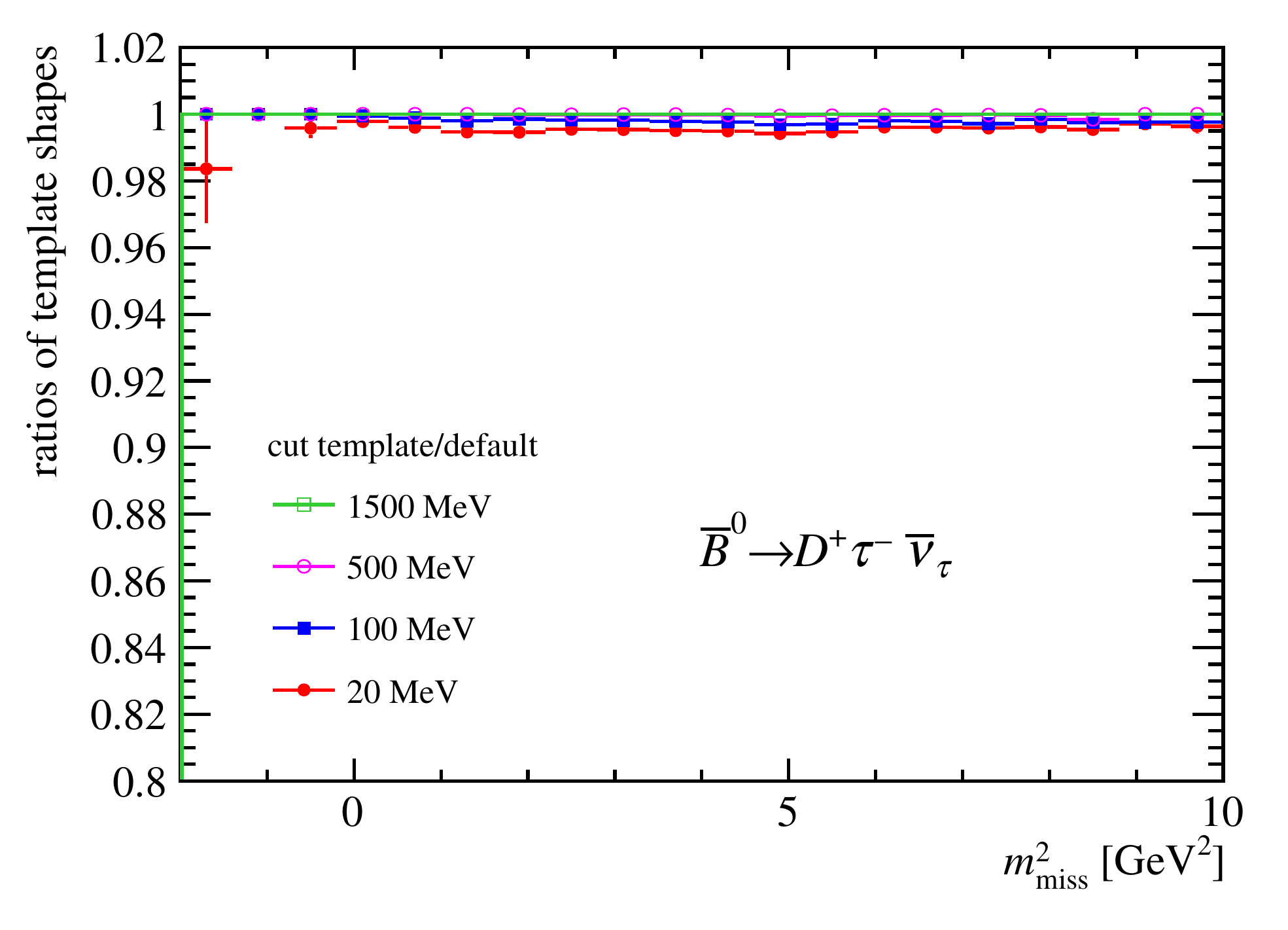}
  \end{center}
  \caption{
   Ratios of the cut \mmiss distribution over the default \mmiss distributions for the  $\bar{B}^0\to D^+\ell^-\bar{\nu_\ell}$ 
  decays, for various \emax cuts. On the left for the \mun and on the right for 
  the \taum mode.}
  \label{fig:mmiss_emaxcuts}
\end{figure*}

When applying cuts on $E_{\gamma}$, the templates shapes change. This is most clearly seen in the distributions of \mmiss, shown for the \Bzb decay
in Fig.~\ref{fig:mmiss_emaxcuts}. Especially in the \mun decay mode the effect is large, altering the shape at high values of \mmiss. Since this feature is not present in the \taum mode, this does not cancel when measuring the ratio \RDp.
For completeness, changes in the shape of \qsq and \emu for the \mun mode are shown in \ref{sec:appendix}.

The number of events generated to simulate data is determined from the estimated number of events
that \lhcb gathered during their Run II data-taking period. The estimate 
takes into account the \B-production cross-section at 13~TeV, branching fractions, and
assumes the average reconstruction efficiency is the same as in Ref.~\cite{Aaij:2015yra}. 
This results in data samples of $1.0\times10^6$ and $0.5\times10^5$ for the 
$\Bzb\to\Dp\ellm\neulb$ decays, and $4.4\times 10^5$ and $2.3\times10^4$ for
the $\Bm\to\Dz\ellm\neulb$ decays, where the first yield represents the \mun sample, and the second the \taum sample. In an actual analysis, the efficiencies for $\B\to\D\ell\neulb$ decays are likely
higher than those for $\B\to\D^{*}\ell\neulb$ decays, where $\D^{*}$ is reconstructed in the $\D\pi$ decay mode.

The measured value of \RD is determined from two components. The first is the ratio of  
reconstruction efficiencies
$\varepsilon_{\mu}$ and $\varepsilon_{\tau}$ for the \mun and \taum samples, respectively, which takes into 
account the selection requirements described earlier in this section. The second component
is the fraction of semitauonic decays in the sample, $f_{\tau}$, determined from the three-dimensional template fit (the absence of background events in the simulated samples implies that the fraction of \mun and \taum components add up to one).
These are combined to measure \RD as
\begin{align}
    \RD = \frac{f_{\tau}}{1-f_{\tau}} \frac{\varepsilon_{\mu}}{\varepsilon_{\tau}} \, .
\end{align}
The exercise of generating pseudo-experiments is repeated 10.000 times after which the spread of the measured values of \RD is taken as the statistical uncertainty.

\begin{figure*}[tb]
  \begin{center}
    \includegraphics[width=0.48\linewidth]{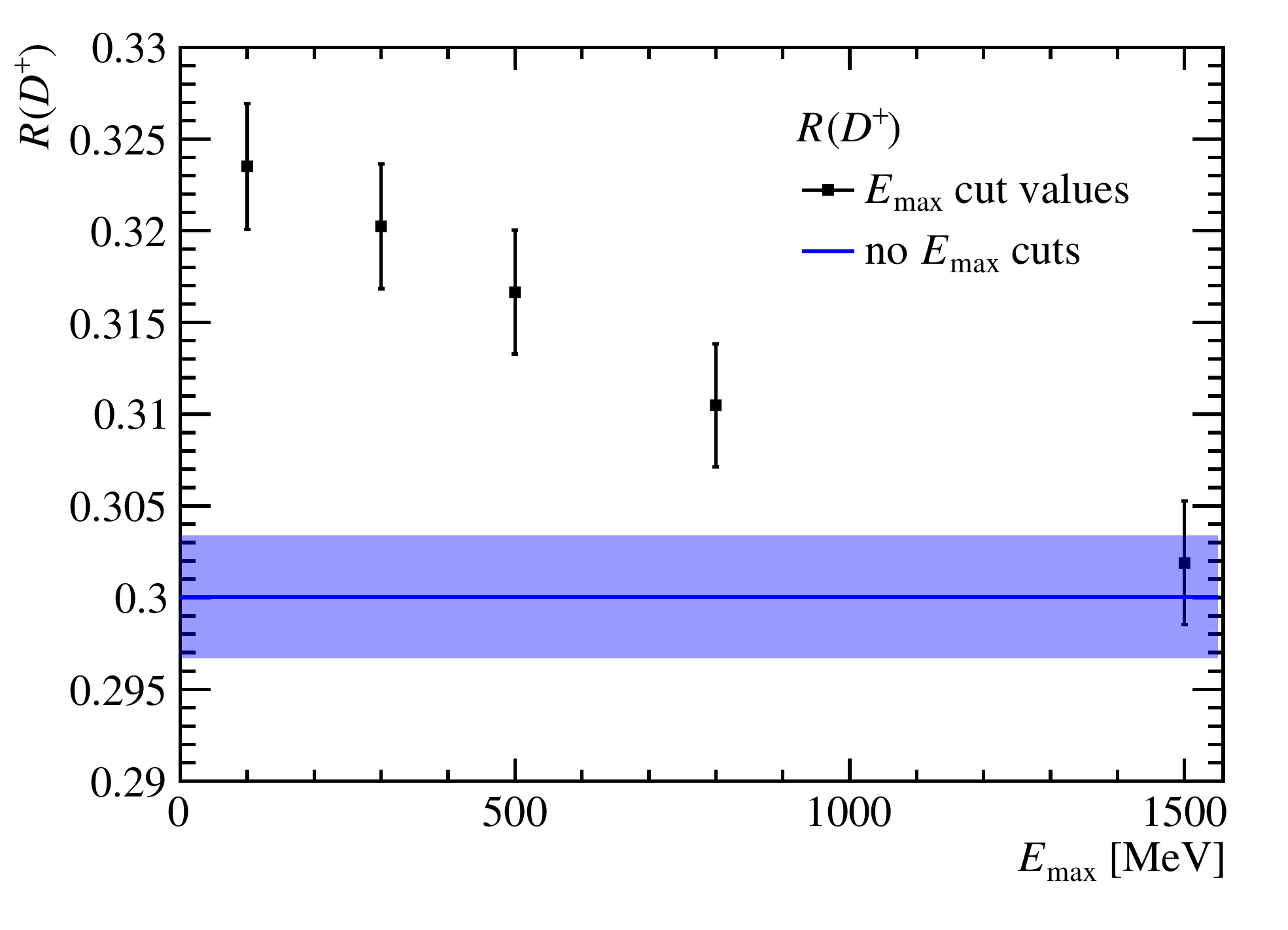}        
    \includegraphics[width=0.48\linewidth]{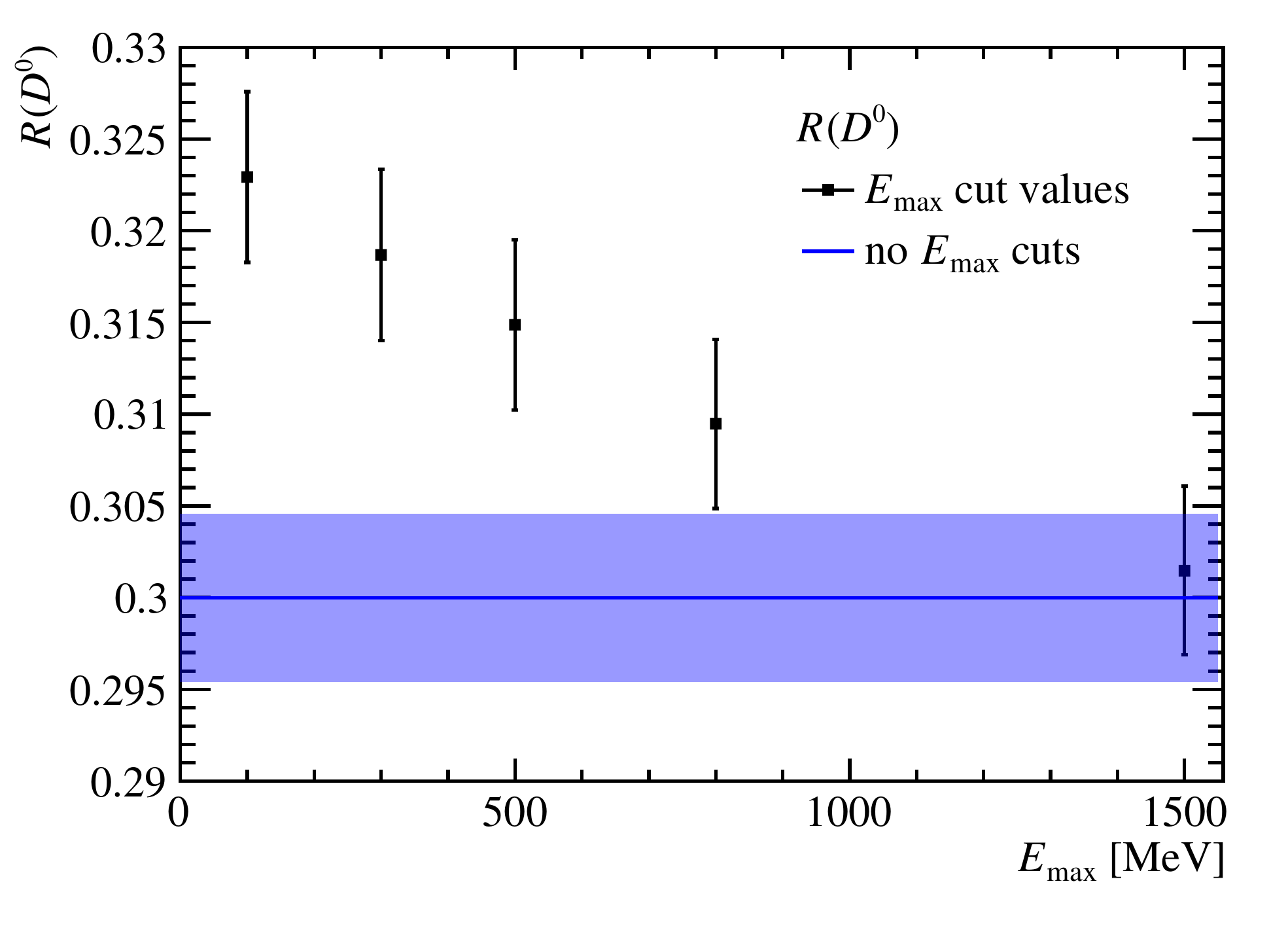}    
  \end{center}
  \caption{Values of \RD as a function of \emax measured in a simplified \lhcb-like analysis of \RDp (left) and \RDz (right).
  The error bars reflect the statistical uncertainties from the generated MC samples. %Specifically, they come by repeating the analysis 10.000 times and taking the spread of the outcomes as the statistical uncertainty. 
  The blue bands correspond to fit results obtained with no \emax cuts. 
  }
  \label{fig:RD_Emax}
\end{figure*}

The resulting values of \RDp and \RDz as a function of \emax are shown in Fig.~\ref{fig:RD_Emax}. From here it is clear that there is a significant effect in underestimating the QED radiative corrections which could be up to 0.02 for both \RDp and \RDz values, corresponding to a relative bias of 7.5\%. The largest contribution to the observed bias is due to the fit fraction $f_{\tau}$, which is strongly affected by the shapes of the \mun templates. Instead, the ratio of efficiencies ${\varepsilon_{\mu}}/{\varepsilon_{\tau}}$ is only marginally dependent on \emax. 
However, this last statement holds only for this specific case study. Different sets of selection cuts or different experimental environments could indeed introduce a significant bias also in the ratio of efficiencies.
The observed bias can be understood when looking at the \mmiss distribution in Fig.~\ref{fig:RF_shapes} and Fig.~\ref{fig:mmiss_emaxcuts}. When cutting on $E_{\gamma}$, part of 
the tail of the \mun distribution is removed, which is compensated by a higher \taum 
fraction in the fit. 

In an actual analysis there are radiative corrections in MC and radiated photons in data. Therefore, it is useful to check the above approach using an alternative strategy. In this case, the templates include all QED corrections predicted by \photos while an \emax cut is applied on the pseudo-experiments. This approach leads to an overestimate on the QED corrections, resulting in a negative bias on the \RD values. The results for \RDp and \RDz as a function of \emax are reported in \ref{sec:appendix}. The corrections are of the exact same size as those in the baseline approach. 

It is worth to note that, despite the fact that \lhcb does not cut explicitly on $E_{\gamma}$ in its analyses, indirect cuts on the total radiated energy are applied through \eg requirements on isolation variables or inefficient reconstruction algorithms for low momentum particles. This could alter the observed bias if $E_{\gamma}$ is not simulated correctly.

These studies show that radiative corrections play a crucial role in \RD measurements.
Since part of these corrections are already simulated in \photos, the above exercise shows the effect 
of a worst-case scenario. Nevertheless, additional effects such as the Coulomb correction, 
as detailed in next section, or the calculation for energies greater than 100~MeV are becoming
fundamental in view of the increased experimental precision expected in the coming years.
Also, these quantitative effects strongly depend on explicit or implicit cuts on radiative photons and must be carefully evaluated for each analysis measuring \RD. 

\subsection{Coulomb correction}
\begin{figure*}[tb]
  \begin{center}
    \includegraphics[width=0.32\linewidth]{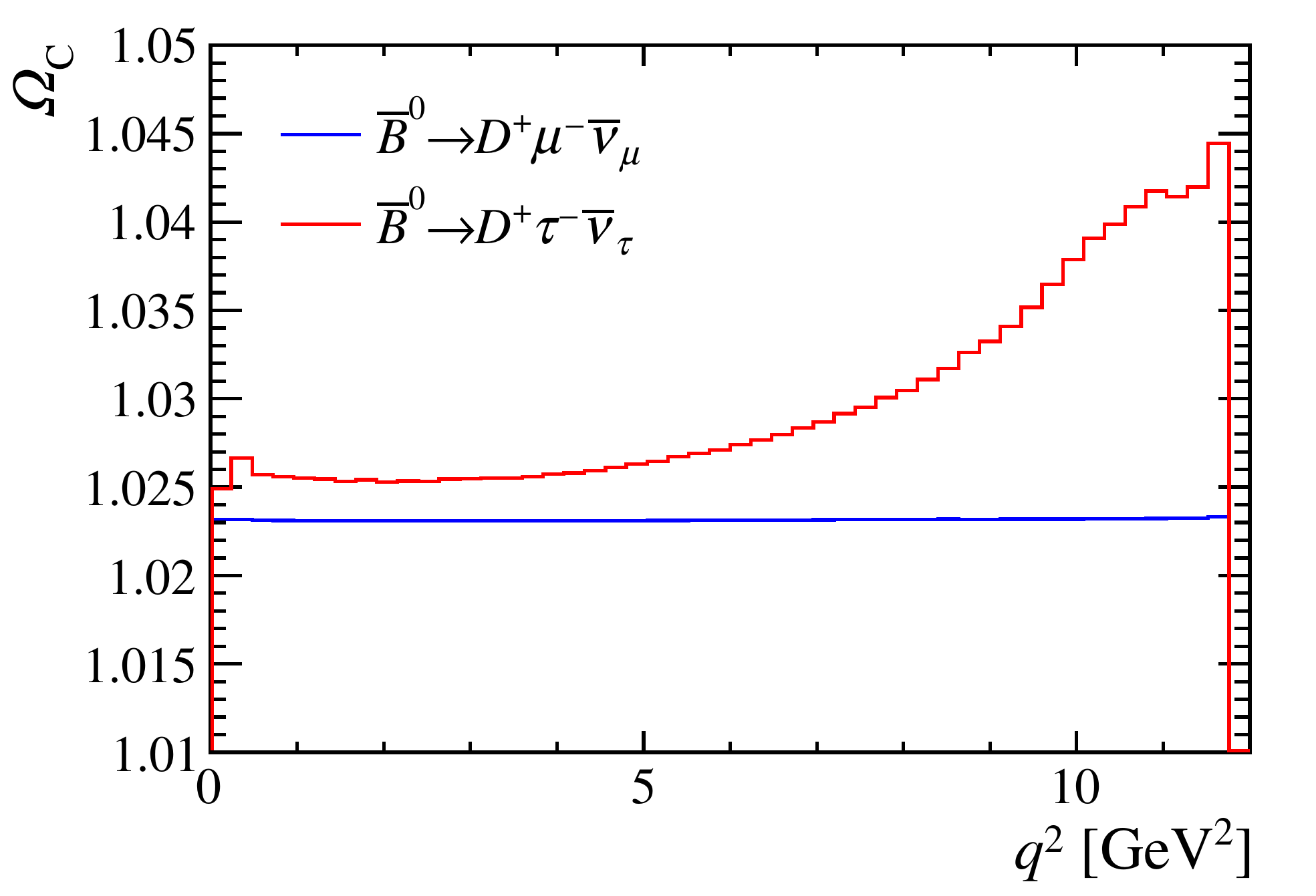}
    \includegraphics[width=0.32\linewidth]{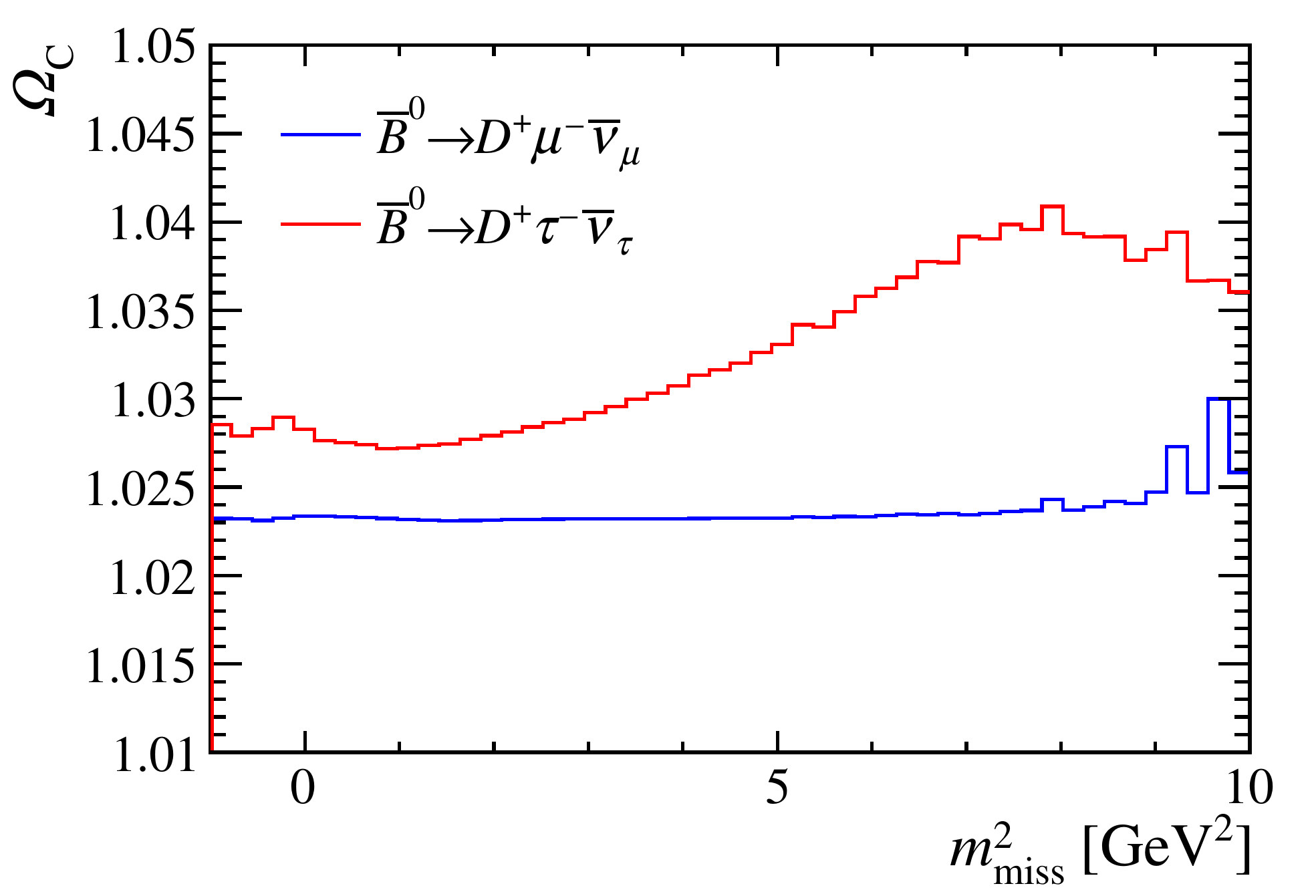}
    \includegraphics[width=0.32\linewidth]{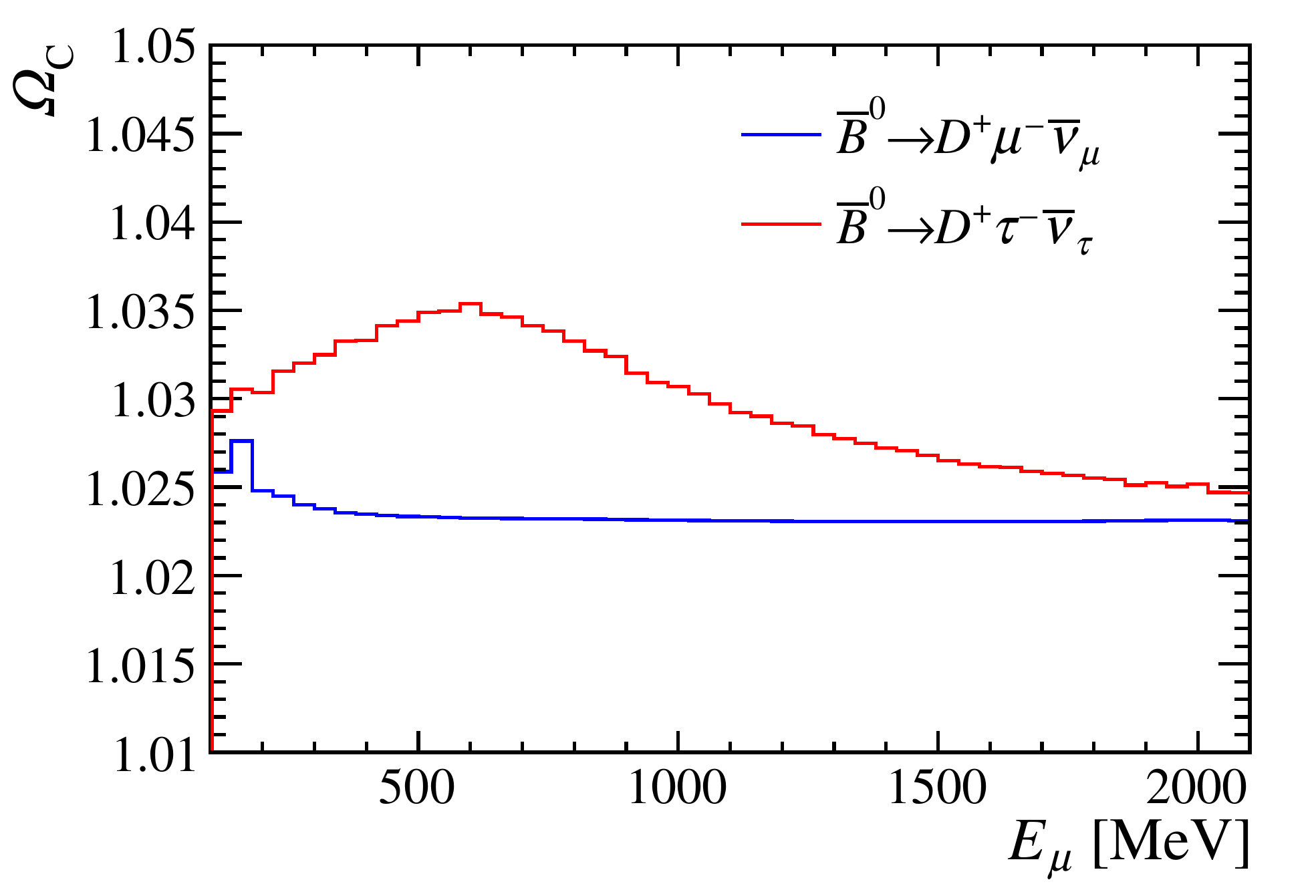}    
  \end{center}
  \caption{
  Coulomb correction as a function of \qsq, \mmiss, and \emu for $\Bzb\to\Dp\ellm\neulb$ decays, where $\ellm = \mun, \taum$.}
  \label{fig:qsquare_Coulomb}
\end{figure*}

Beyond affecting the SM prediction of \RDp, the Coulomb correction impacts the experimental results by changing the shape of the fit templates.
This is evaluated in the $\Bzb\to\Dp\ellm\neulb$ decay by weighting each event by the term \OmegaC.\footnote{The contribution due to the non-factorizable loop corrections to the tree-level differential decay rate, called $\tilde{\Gamma}^{\Dp}$ in Ref.\cite{deBoer:2018ipi}, is small and not implemented in \photos. For this reason, the Coulomb correction \OmegaC can be introduced in \photos as a global factor to the uncorrected differential decay rate.} The changes in the shape of the \qsq, \mmiss and \emu distributions are shown in Fig.~\ref{fig:qsquare_Coulomb}.
While for the \mun mode
\OmegaC is mostly constant, for the \taum mode there is a dependence on each of the three variables due to the smaller relative velocity. 
To quantify the effect of the Coulomb correction, the above analysis is repeated, without including any \emax cuts. The Coulomb correction is applied to the pseudo-experiments, but not to the fit templates, resulting in a relative shift of about -1.0\% on \RDp. This effect can even be amplified by selecting certain regions of phase space.

\vspace{-0.2cm}

\section{Conclusions and recommendations}
\label{sec:Conclusions}

The work in Ref.~\cite{deBoer:2018ipi} describes QED corrections which are not fully 
included in \photos. These corrections affect the semimuonic and semitauonic modes differently
at the level of a few percent. Ignoring the Coulomb correction, there is more radiated energy in the 
calculation in Ref.~\cite{deBoer:2018ipi} than in \photos for the \Bzb decays, while this is the other way around for 
the \Bm decays. In the ratio \RD, this small discrepancy mostly cancels out. 
However, the main difference between the QED corrections on \RDp and \RDz, which is up to 1\%, is due to the Coulomb correction that only affects \RDp.

Coulomb interactions
are not simulated by \photos and mainly affect the kinematics of semitauonic decays, which in turn 
influence the shape of distributions used to determine the signal yields in \lhcb, 
\babar, and \belle analyses. These effects can alter values of \RD up to 1\% in an 
\lhcb-like analysis, and should be evaluated precisely for each measurement. 

Using a simplified \lhcb-like analysis, it is shown that over- or underestimating radiative corrections could
bias measurements of \RD up to 7\% in an extreme case. This results in a bias of 0.02 
on the value of \RD, and should be studied further when performing these types of measurements, including a realistic evaluation of cuts on $E_{\gamma}$. 
These effects could potentially be enhanced in measurements from \belle~II \cite{Kou:2018nap} where the resolution on the kinematic variables is better than at \lhcb.

When measuring values of \RD with higher precision, additional calculations of QED 
corrections for $\B\to\D\ell\neul$ decays are necessary. Especially calculations involving high-energy and structure-dependent
photons are currently mostly missing. 

\vspace{-0.3cm}
\begin{acknowledgements}
We are grateful to S. de Boer, T. Kitahara, and I. Nisandzic for the fruitful collaboration, and to U. Egede for his thoughtful comments.
In addition, we thank the Semileptonics \B decays working group of the \lhcb collaboration, and in particular M. De Cian and L. Grillo, for 
their useful feedback throughout the development of this paper. Finally, we thank Z. Was for his helpful insights into the \photos package.
\end{acknowledgements}

\appendix
\section{Additional Plots}
\label{sec:appendix}
\begin{figure*}[tb]
  \begin{center}
    \includegraphics[width=0.48\linewidth]{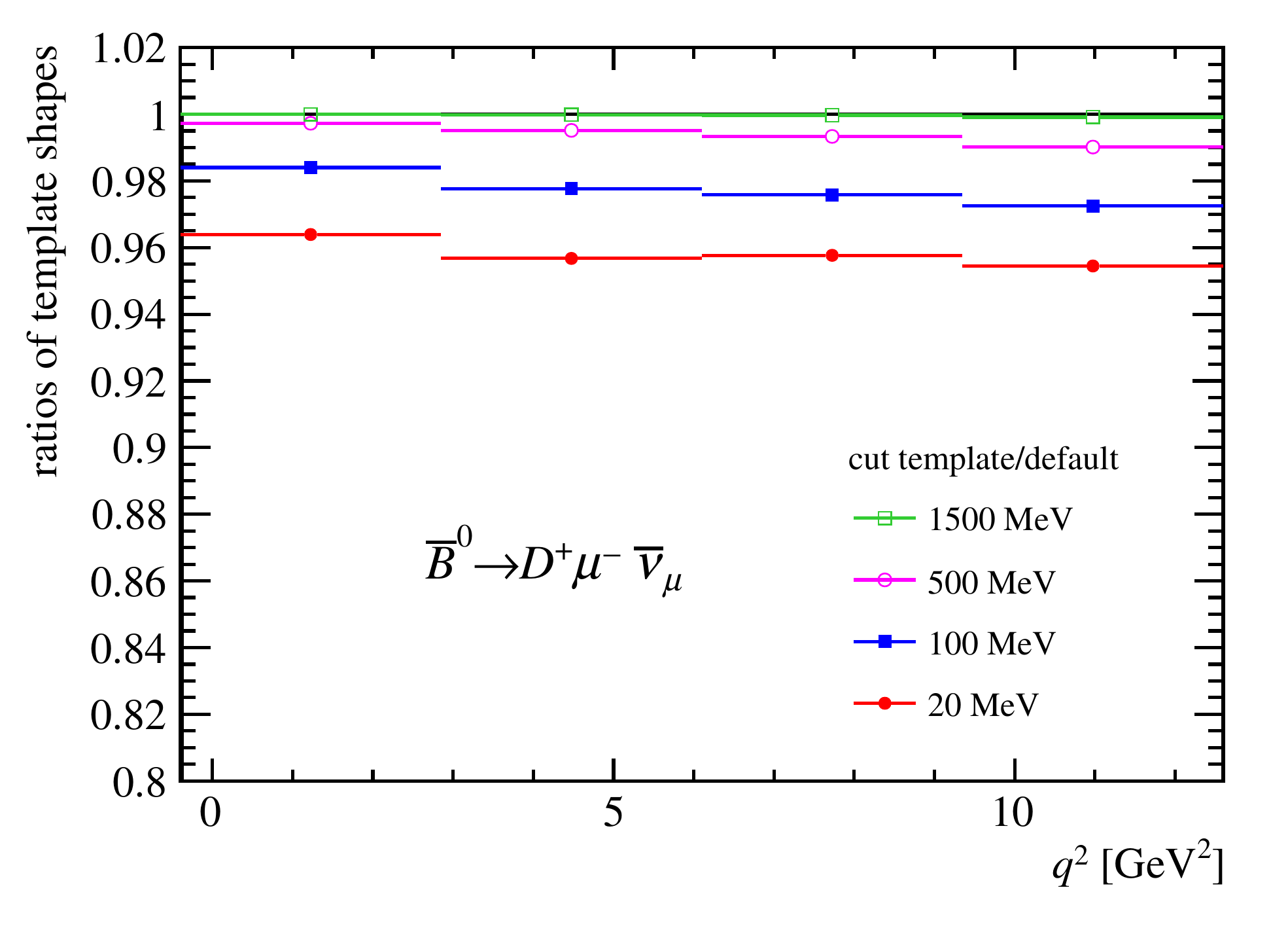}
    \includegraphics[width=0.48\linewidth]{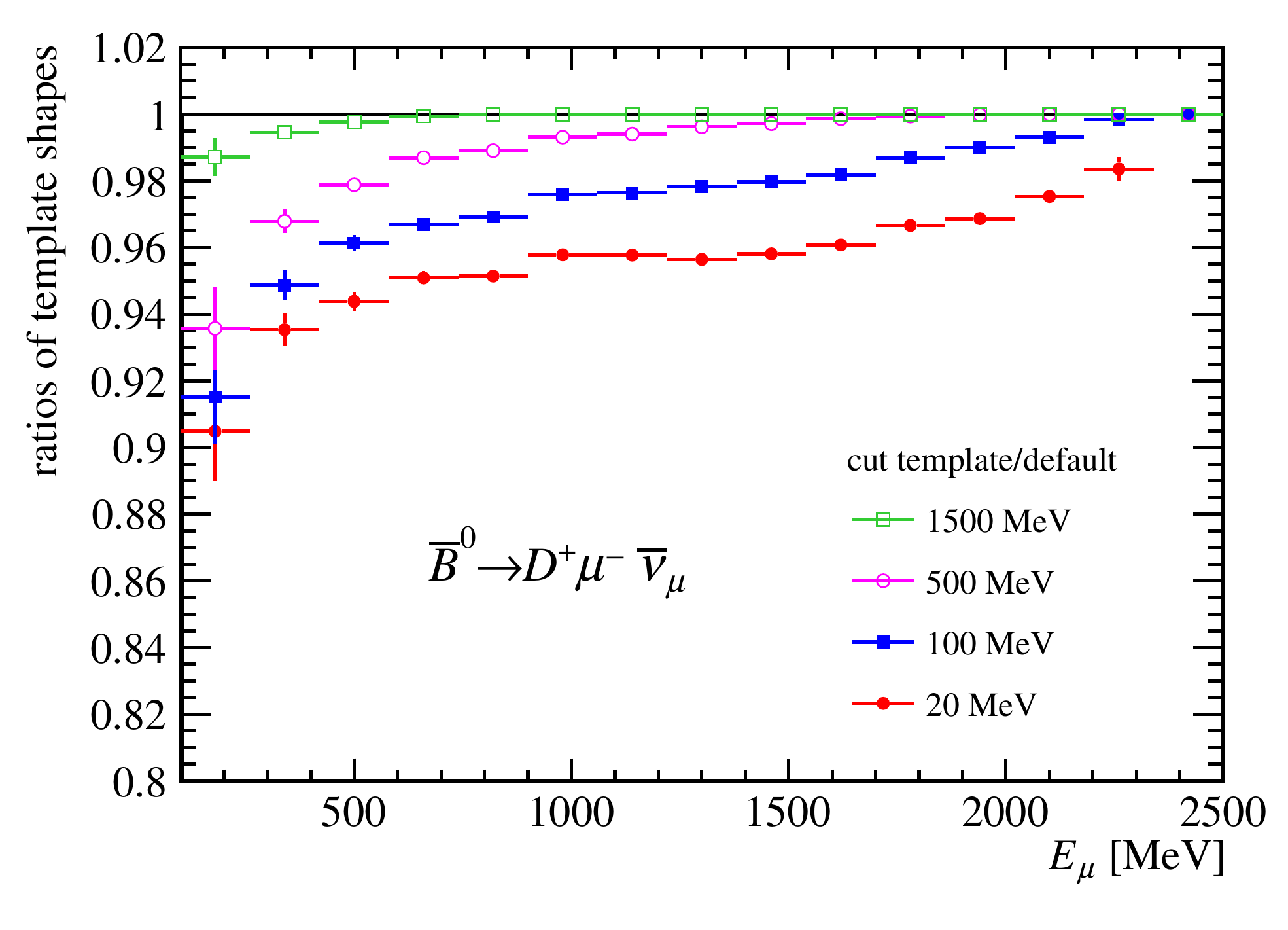}
  \end{center}
  \caption{Ratios of the cut \qsq (left) and \emu (right) distributions over the corresponding default distributions for $\Bzb\to\Dp\mun\neumb$ decays, for various \emax cuts.}
  \label{fig:q2_El_emaxcuts}
\end{figure*}

The fits on the pseudo-experiments are performed on the three variables \emu, \mmiss and \qsq, as described in Sect.~\ref{sec:dummyAnalysis}. The effect of cutting on \emax on the shape of the \emu and \mmiss templates is shown in Fig.~\ref{fig:q2_El_emaxcuts} for the $\Bzb\to\Dp\mun\neumb$ decay. Analogous plots for the semitauonic mode show a negligible dependence on the \emax cut.

\begin{figure*}[tb]
  \begin{center}
    \includegraphics[width=0.48\linewidth]{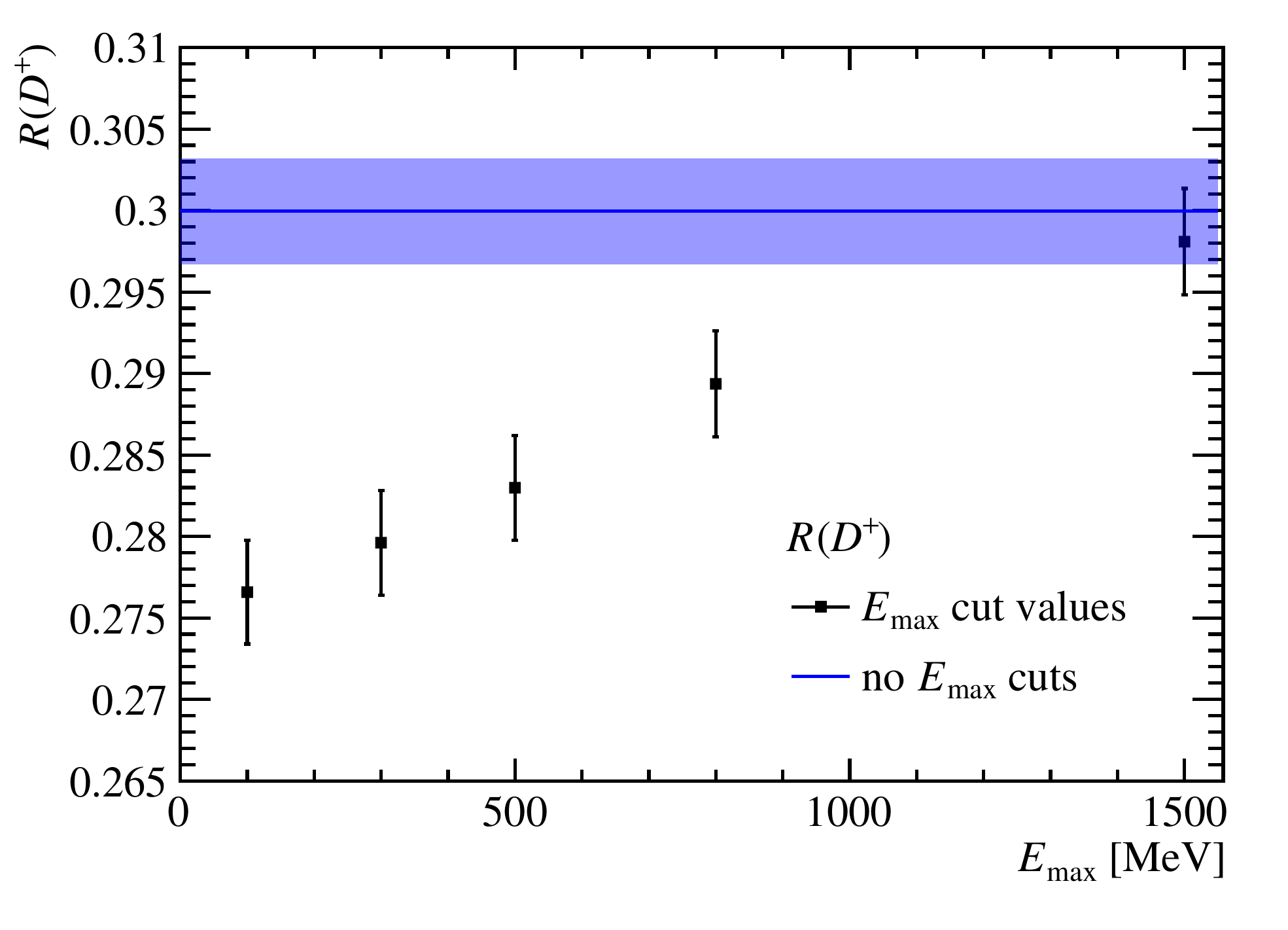}        
    \includegraphics[width=0.48\linewidth]{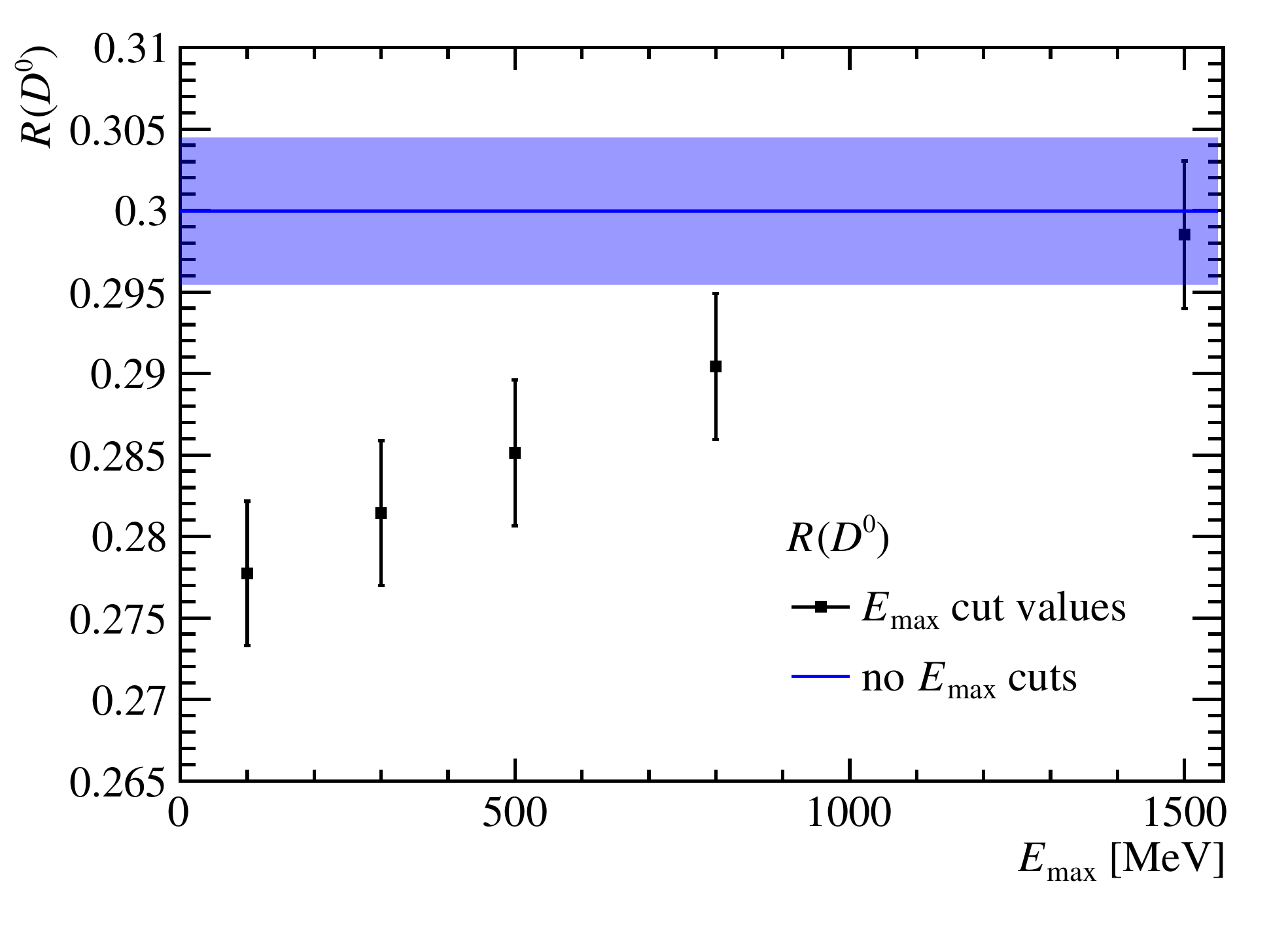}    
  \end{center}
  \caption{Values of \RD as a function of \emax from a simplified \lhcb-like analysis of \RDp (left) and \RDz (right). In this alternative approach, templates are generated according to \photos predictions and \emax cuts are applied to the pseudo-experiments. The error bars reflect the statistical uncertainties from the generated MC samples. The blue bands correspond to fit results obtained with no \emax cuts.
  }
  \label{fig:RD_plots_strategy2}
\end{figure*}

The results of performing the simplified LHCb-like analysis with the alternative strategy are shown in Fig.~\ref{fig:RD_plots_strategy2}. These results are obtained using templates with an $E_{\gamma}$ distribution in agreement with \photos predictions, and pseudo-experiments with cuts on $E_{\gamma}$ applied.  

\setboolean{inbibliography}{true}
\bibliographystyle{LHCb}
\bibliography{main}

\end{document}